\documentclass[11pt]{article}
\usepackage{hyperref,url}
\usepackage[utf8x]{inputenc} 
\usepackage{latexsym,graphicx,mathrsfs,amsfonts,subfig}
\usepackage{slashed, color}
\usepackage{multirow}
\usepackage{tikz}
\usepackage{empheq}
\usepackage{amsmath}
\allowdisplaybreaks[2]
\usepackage[multiple]{footmisc}
\usepackage{textcomp}
\usepackage{amssymb}
\usetikzlibrary{positioning}
\usepackage{amscd}
\usepackage{amsthm}
\usepackage{array} 

\renewcommand{\thefootnote}{\fnsymbol{footnote}}
\numberwithin{equation}{section}
\usepackage{dsfont}
\usepackage[toc,page]{appendix}

\DeclareFontFamily{U}{MnSymbolC}{}
\DeclareSymbolFont{MnSyC}{U}{MnSymbolC}{m}{n}
\DeclareFontShape{U}{MnSymbolC}{m}{n}{
	<-6>  MnSymbolC5
	<6-7>  MnSymbolC6
	<7-8>  MnSymbolC7
	<8-9>  MnSymbolC8
	<9-10> MnSymbolC9
	<10-12> MnSymbolC10
	<12->   MnSymbolC12}{}
\DeclareMathSymbol{\intprod}{\mathbin}{MnSyC}{'270}
\newcommand{\ov}{\overline}
\newcommand{\C}{\mathbb{C}}

\newcommand{\R}{\mathbb{R}}

\newcommand{\til}{\widetilde}

\let\nc\newcommand
\let\renc\renewcommand
\nc{\wbar}{\overline}
\let\td\tilde
\let\wtd\widetilde
\let\wht\widehat
\let\mcl\mathcal

\nc{\ab}{{\bar{a}}} \nc{\at}{\tilde{a}} \nc{\ah}{\hat{a}}
\nc{\bb}{{\bar{b}}} 
\nc{\bh}{\hat{b}}
\nc{\cb}{{\bar{c}}} \nc{\ct}{\tilde{c}} 
\nc{\db}{{\bar{d}}} \nc{\dt}{\tilde{d}} \renc{\dh}{\hat{d}}
\nc{\eb}{{\bar{e}}} \nc{\et}{\tilde{e}} \nc{\eh}{\hat{e}}
\nc{\fb}{{\bar{f}}} \nc{\ft}{\tilde{f}} \nc{\fh}{\hat{f}}
\nc{\ib}{{\bar{\imath}}} \nc{\ih}{\hat{\imath}} 
\nc{\jb}{{\bar{\jmath}}} \nc{\jt}{\tilde{\jmath}} \nc{\jh}{\hat{\jmath}}
\nc{\kb}{{\bar{k}}} \nc{\kt}{\tilde{k}} \nc{\kh}{\hat{k}}
\nc{\lb}{{\bar{l}}} \nc{\lt}{\tilde{l}} \nc{\lh}{\hat{l}}
\nc{\mb}{{\bar{m}}} \nc{\mt}{\tilde{m}} \nc{\mh}{\hat{m}}
\nc{\nb}{{\bar{n}}} \nc{\nt}{\tilde{n}} \nc{\nh}{\hat{n}}
\nc{\ob}{{\bar{o}}} \nc{\ot}{\tilde{o}} \nc{\oh}{\hat{o}}
\nc{\pb}{{\bar{p}}} \nc{\pt}{\tilde{p}} \nc{\ph}{\hat{p}}
\nc{\qb}{{\bar{q}}} \nc{\qt}{\tilde{q}} \nc{\qh}{\hat{q}}
\nc{\rb}{{\bar{r}}} \nc{\rt}{\tilde{r}} \nc{\rh}{\hat{r}}
\renc{\sb}{{\bar{s}}} \nc{\st}{\tilde{s}} \nc{\sh}{\hat{s}}
\nc{\tb}{{\bar{t}}} \renc{\th}{\hat{t}} 
\nc{\ub}{{\bar{u}}} \nc{\ut}{\tilde{u}} \nc{\uh}{\hat{u}}
\nc{\vb}{{\bar{v}}} \nc{\vt}{\tilde{v}} \nc{\vh}{\hat{v}}
\nc{\wt}{\tilde{w}} \nc{\wh}{\hat{w}}
\nc{\xb}{{\bar{x}}} \nc{\xt}{\tilde{x}} \nc{\xh}{\hat{x}}
\nc{\yb}{{\bar{y}}} \nc{\yt}{\tilde{y}} \nc{\yh}{\hat{y}}
\nc{\zb}{{\bar{z}}} \nc{\zt}{\tilde{z}} 

\nc{\Ab}{\wbar{A}} \nc{\At}{\wtd{A}} \nc{\Ah}{\wht{A}}
\nc{\Bb}{\wbar{B}} \nc{\Bt}{\wtd{B}} \nc{\Bh}{\wht{B}}
\nc{\Cb}{\wbar{C}} \nc{\Ct}{\wtd{C}} \nc{\Ch}{\wht{C}}
\nc{\Db}{\wbar{D}} \nc{\Dt}{\wtd{D}} \nc{\Dh}{\wht{D}}
\nc{\Eb}{\wbar{E}} \nc{\Et}{\wtd{E}} \nc{\Eh}{\wht{E}}
\nc{\Fb}{\wbar{F}} \nc{\Ft}{\wtd{F}} \nc{\Fh}{\wht{F}}
\nc{\Gb}{\wbar{G}} \nc{\Gt}{\wtd{G}} \nc{\Gh}{\wht{G}}
\nc{\Hb}{\wbar{H}} \nc{\Ht}{\wtd{H}} \nc{\Hh}{\wht{H}}
\nc{\Ib}{\wbar{I}} \nc{\It}{\wtd{I}} \nc{\Ih}{\wht{I}}
\nc{\Jb}{\wbar{J}} \nc{\Jt}{\wtd{J}} \nc{\Jh}{\wht{J}}
\nc{\Kb}{\wbar{K}} \nc{\Kt}{\wtd{K}} \nc{\Kh}{\wht{K}}
\nc{\Lb}{\wbar{L}} \nc{\Lt}{\wtd{L}} \nc{\Lh}{\wht{L}}
\nc{\Mb}{\wbar{M}} \nc{\Mt}{\wtd{M}} \nc{\Mh}{\wht{M}}
\nc{\Nb}{\wbar{N}} \nc{\Nt}{\wtd{N}} \nc{\Nh}{\wht{N}}
\nc{\Ob}{\wbar{O}} \nc{\Ot}{\wtd{O}} \nc{\Oh}{\wht{O}}
\nc{\Pb}{\wbar{P}} \nc{\Pt}{\wtd{P}} \nc{\Ph}{\wht{P}}
\nc{\Qb}{\wbar{Q}} \nc{\Qt}{\wtd{Q}} \nc{\Qh}{\wht{Q}}
\nc{\Rb}{\wbar{R}} \nc{\Rt}{\wtd{R}} \nc{\Rh}{\wht{R}}
\nc{\Sb}{\wbar{S}} \nc{\St}{\wtd{S}} \nc{\Sh}{\wht{S}}
\nc{\Tb}{\wbar{T}} \nc{\Tt}{\wtd{T}} \nc{\Th}{\wht{T}}
\nc{\Ub}{\wbar{U}} \nc{\Ut}{\wtd{U}} \nc{\Uh}{\wht{U}}
\nc{\Vb}{\wbar{V}} \nc{\Vt}{\wtd{V}} \nc{\Vh}{\wht{V}}
\nc{\Wb}{\wbar{W}} \nc{\Wt}{\wtd{W}} \nc{\Wh}{\wht{W}}
\nc{\Xb}{\wbar{X}} \nc{\Xt}{\wtd{X}} \nc{\Xh}{\wht{X}}
\nc{\Yb}{\wbar{Y}} \nc{\Yt}{\wtd{Y}} \nc{\Yh}{\wht{Y}}
\nc{\Zb}{\wbar{Z}} \nc{\Zt}{\wtd{Z}} \nc{\Zh}{\wht{Z}}

\nc{\CA}{\mcl{A}} \nc{\CAb}{\wbar{\CA}} \nc{\CAt}{\wtd{\CA}} \nc{\CAh}{\wht{\CA}}
\nc{\CB}{\mcl{B}} \nc{\CBb}{\wbar{\CB}} \nc{\CBt}{\wtd{\CB}} \nc{\CBh}{\wht{\CB}}
\nc{\CC}{\mcl{C}} \nc{\CCb}{\wbar{\CC}} \nc{\CCt}{\wtd{\CC}} \nc{\CCh}{\wht{\CC}}
\nc{\cDt}{\wtd{\cC}} \nc{\cDh}{\wht{\cD}}
\nc{\CE}{\mcl{E}} \nc{\CEb}{\wbar{\CE}} \nc{\CEt}{\wtd{\CE}} \nc{\CEh}{\wht{\CE}}
\nc{\CF}{\mcl{F}} \nc{\CFb}{\wbar{\CF}} \nc{\CFt}{\wtd{\CF}} \nc{\CFh}{\wht{\CF}}
\nc{\CG}{\mcl{G}} \nc{\CGb}{\wbar{\CG}} \nc{\CGt}{\wtd{\CG}} \nc{\CGh}{\wht{\CG}}
\nc{\CH}{\mcl{H}} \nc{\CHb}{\wbar{\CH}} \nc{\CHt}{\wtd{\CH}} \nc{\CHh}{\wht{\CH}}
\nc{\CI}{\mcl{I}} \nc{\CIb}{\wbar{\CI}} \nc{\CIt}{\wtd{\CI}} \nc{\CIh}{\wht{\CI}}
\nc{\CJ}{\mcl{J}} \nc{\CJb}{\wbar{\CJ}} \nc{\CJt}{\wtd{\CJ}} \nc{\CJh}{\wht{\CJ}}
\nc{\CK}{\mcl{K}} \nc{\CKb}{\wbar{\CK}} \nc{\CKt}{\wtd{\CK}} \nc{\CKh}{\wht{\CK}}
\nc{\CL}{\mcl{L}} \nc{\CLb}{\wbar{\CL}} \nc{\CLt}{\wtd{\CL}} \nc{\CLh}{\wht{\CL}}
\nc{\CM}{\mcl{M}} \nc{\CMb}{\wbar{\CM}} \nc{\CMt}{\wtd{\CM}} \nc{\CMh}{\wht{\CM}}
\nc{\CN}{\mcl{N}} \nc{\CNb}{\wbar{\CN}} \nc{\CNt}{\wtd{\CN}} \nc{\CNh}{\wht{\CN}}
\nc{\CO}{\mcl{O}} \nc{\COb}{\wbar{\CO}} \nc{\COt}{\wtd{\CO}} \nc{\COh}{\wht{\CO}}
\nc{\CQ}{\mcl{Q}} \nc{\CQb}{\wbar{\CQ}} \nc{\CQt}{\wtd{\CQ}} \nc{\CQh}{\wht{\CQ}}
\nc{\CR}{\mcl{R}} \nc{\CRb}{\wbar{\CR}} \nc{\CRt}{\wtd{\CR}} \nc{\CRh}{\wht{\CR}}
\nc{\CS}{\mcl{S}} \nc{\CSb}{\wbar{\CS}} \nc{\CSt}{\wtd{\CS}} \nc{\CSh}{\wht{\CS}}
\nc{\CT}{\mcl{T}} \nc{\CTb}{\wbar{\CT}} \nc{\CTt}{\wtd{\CT}} \nc{\CTh}{\wht{\CT}}
\nc{\CU}{\mcl{U}} \nc{\CUb}{\wbar{\CU}} \nc{\CUt}{\wtd{\CU}} \nc{\CUh}{\wht{\CU}}
\nc{\CV}{\mcl{V}} \nc{\CVb}{\wbar{\CV}} \nc{\CVt}{\wtd{\CV}} \nc{\CVh}{\wht{\CV}}
\nc{\CW}{\mcl{W}} \nc{\CWb}{\wbar{\CW}} \nc{\CWt}{\wtd{\CW}} \nc{\CWh}{\wht{\CW}}
\nc{\CX}{\mcl{X}} \nc{\CXb}{\wbar{\CX}} \nc{\CXt}{\wtd{\CX}} \nc{\CXh}{\wht{\CX}}
\nc{\CY}{\mcl{Y}} \nc{\CYb}{\wbar{\CY}} \nc{\CYt}{\wtd{\CY}} \nc{\CYh}{\wht{\CY}}
\nc{\CZ}{\mcl{Z}} \nc{\CZb}{\wbar{\CZ}} \nc{\CZt}{\wtd{\CZ}} \nc{\CZh}{\wht{\CZ}}

\let\eps\epsilon
\let\ups\upsilon
\let\veps\varepsilon
\let\vtht\vartheta
\let\vsgm\varsigma
\let\vphi\varphi
\let\vrho\varrho

\nc{\alphab}{\bar{\alpha}} \nc{\alphat}{\td{\alpha}} \nc{\alphah}{\hat{\alpha}}
\nc{\betab}{\bar{\beta}}   \nc{\betat}{\td{\beta}}   \nc{\betah}{\hat{\beta}} 
\nc{\gammab}{\bar{\gamma}} \nc{\gammat}{\td{\gamma}} \nc{\gammah}{\hat{\gamma}} 
\nc{\deltab}{\bar{\delta}} \nc{\deltat}{\td{\delta}} \nc{\deltah}{\hat{\delta}} 
\nc{\epsilonb}{\bar{\eps}} \nc{\epsilont}{\td{\eps}} \nc{\epsilonh}{\hat{\eps}} 
\nc{\vepsb}{\bar{\veps}}   \nc{\vepst}{\td{\veps}}   \nc{\vepsh}{\hat{\veps}} 
\nc{\zetab}{\bar{\zeta}}   \nc{\zetat}{\td{\zeta}}   \nc{\zetah}{\hat{\zeta}} 
\nc{\etab}{\bar{\eta}}     
\nc{\etah}{\hat{\eta}} 
\nc{\thetab}{\bar{\theta}} \nc{\thetat}{\td{\theta}} \nc{\thetah}{\hat{\theta}} 
\nc{\vthetab}{\bar{\vtht}} \nc{\vthetat}{\td{\vtht}} \nc{\vthetah}{\hat{\vtht}} 
\nc{\lambdat}{\td{\lambda}} \nc{\lambdah}{\hat{\lambda}} 
\nc{\iotab}{\bar{\iota}}   \nc{\iotat}{\td{\iota}}   \nc{\iotah}{\hat{\iota}} 
\nc{\kappab}{\bar{\kappa}} \nc{\kappat}{\td{\kappa}} \nc{\kappah}{\hat{\kappa}} 
\nc{\lmdb}{\bar{\lmd}}     \nc{\lmdt}{\td{\lmd}}     \nc{\lmdh}{\hat{\lmd}} 
\nc{\mub}{\bar{\mu}}       \nc{\mut}{\td{\mu}}       \nc{\muh}{\hat{\mu}} 
\nc{\nub}{\bar{\nu}}       \nc{\nut}{\td{\nu}}       \nc{\nuh}{\hat{\nu}} 
\nc{\xib}{\bar{\xi}}       \nc{\xit}{\td{\xi}}       \nc{\xih}{\hat{\xi}} 
\nc{\pib}{\bar{\pi}}       \nc{\pit}{\td{\pi}}       \nc{\pih}{\hat{\pi}} 
\nc{\vpib}{\bar{\vpi}}     \nc{\vpit}{\td{\vpi}}     \nc{\vpih}{\hat{\vpi}} 
\nc{\rhob}{\bar{\rho}}     \nc{\rhot}{\td{\rho}}     \nc{\rhoh}{\hat{\rho}} 
\nc{\vrhob}{\bar{\vrho}}   \nc{\vrhot}{\td{\vrho}}   \nc{\vrhoh}{\hat{\vrho}} 
\nc{\sigmab}{\bar{\sigma}} \nc{\sigmat}{\td{\sigma}} \nc{\sigmah}{\hat{\sigma}} 
\nc{\vsigmab}{\bar{\vsgm}} \nc{\vsigmat}{\td{\vsgm}} \nc{\vsigmah}{\hat{\vsgm}} 
\nc{\taub}{\bar{\tau}}     \nc{\taut}{\td{\tau}}     \nc{\tauh}{\hat{\tau}} 
\nc{\upsb}{\bar{\ups}} \nc{\upst}{\td{\ups}} \nc{\upsh}{\hat{\ups}} 
\nc{\phib}{\bar{\phi}}     \nc{\phit}{\td{\phi}}     \nc{\phih}{\hat{\phi}} 
\nc{\varphib}{\bar{\vphi}}   \nc{\varphit}{\td{\vphi}}   \nc{\varphih}{\hat{\vphi}} 
\nc{\chib}{\bar{\chi}}     
\nc{\chih}{\hat{\chi}} 
\nc{\psib}{\bar{\psi}}     
\nc{\psih}{\hat{\psi}} 
\nc{\omegab}{\bar{\omega}} \nc{\omegat}{\td{\omega}} \nc{\omegah}{\hat{\omega}} 

\nc{\Gammab}{\wbar{\Gamma}}     \nc{\Gammat}{\wtd{\Gamma}}     \nc{\Gammah}{\wht{\Gamma}}
\nc{\Deltab}{\wbar{\Delta}}     \nc{\Deltat}{\wtd{\Delta}}     \nc{\Deltah}{\wht{\Delta}}
\nc{\Thetab}{\wbar{\Theta}}     \nc{\Thetat}{\wtd{\Theta}}     \nc{\Thetah}{\wht{\Theta}}
\nc{\Lambdab}{\wbar{\Lambda}}   \nc{\Lambdat}{\wtd{\Lambda}}   \nc{\Lambdah}{\wht{\Lambda}}
\nc{\Xib}{\wbar{\Xi}}           \nc{\Xit}{\wtd{\Xi}}           \nc{\Xih}{\wht{\Xi}}
\nc{\Pib}{\wbar{\Pi}}           \nc{\Pit}{\wtd{\Pi}}           \nc{\Pih}{\wht{\Pi}}
\nc{\Sigmab}{\wbar{\Sigma}}     \nc{\Sigmat}{\wtd{\Sigma}}     \nc{\Sigmah}{\wht{\Sigma}}
\nc{\Upsilonb}{\wbar{\Upsilon}} \nc{\Upsilont}{\wtd{\Upsilon}} \nc{\Upsilonh}{\wht{\Upsilon}}
\nc{\Phib}{\wbar{\Phi}}         \nc{\Phit}{\wtd{\Phi}}         \nc{\Phih}{\wht{\Phi}}
\nc{\Psib}{\wbar{\Psi}}         \nc{\Psit}{\wtd{\Psi}}         \nc{\Psih}{\wht{\Psi}}
\nc{\Omegab}{\wbar{\Omega}}     \nc{\Omegat}{\wtd{\Omega}}     \nc{\Omegah}{\wht{\Omega}}
\nc{\Varepsilon}{\mathcal{E}}

\newcommand{\cA}{{\cal A}}

\newcommand{\fA}{\mathfrak A}

\newcommand{\Tr}{{\textrm{Tr}\;}}

\newcommand{\cF}{{\cal F}}

\newcommand{\quar}{\frac{1}{4}}

\newcommand{\cD}{{\cal D}}

\newcommand{\si}{\sigma}

\newcommand{\nn}{\newline}

\nc{\balpha}{\bar{\alpha}}
\nc{\bbeta}{\bar{\beta}}
\nc{\bgamma}{\bar{\gamma}}
\nc{\bm}{\bar{m}}
\nc{\bn}{\bar{n}}
\nc{\bp}{\bar{p}}
\nc{\al}{\alpha}
\nc{\bt}{\beta}
\nc{\gm}{\gamma}
\nc{\zh}{\wht{z}}
\nc{\zhb}{\ov{\wht{z}}}
\nc{\mbh}{\wht{\ov{m}}}
\nc{\bc}{|_{x^2=0}}

\nc{\tal}{\til{\al}}
\nc{\tbt}{\til{\bt}}
\nc{\tgm}{\til{\gm}}

\nc{\wb}{\ov{w}}
\nc{\teta}{\til{\eta}}
\nc{\tpsi}{\til{\psi}}

\def\IL{\relax{\rm I\kern-.18em L}}
\def\IH{\relax{\rm I\kern-.18em H}}
\def\IB{\relax{\rm I\kern-.18em B}}
\def\ID{\relax{\rm I\kern-.18em D}}
\def\IE{\relax{\rm I\kern-.18em E}}
\def\IF{\relax{\rm I\kern-.18em F}}

\def\IG{\relax\hbox{$\inbar\kern-.3em{\rm G}$}}
\def\IGa{\relax\hbox{${\rm I}\kern-.18em\Gamma$}}
\def\IH{\relax{\rm I\kern-.18em H}}
\def\II{\relax{\rm I\kern-.18em I}}
\def\IK{\relax{\rm I\kern-.18em K}}
\def\IP{\relax{\rm I\kern-.18em P}}
\def\IQ{\relax\hbox{$\inbar\kern-.3em{\rm Q}$}}

\def\hat{\widehat}
\def\CM {{\cal M}}
\def\CN {{\cal N}}
\def\CR {{\cal R}}

\def\CF {{\cal F}}
\def\CJ {{\cal J}}

\def\CL {{\cal L}}
\def\CV {{\cal V}}
\def\CO {{\cal O}}
\def\CZ {{\cal Z}}
\def\CE {{\cal E}}
\def\CG {{\cal G}}
\def\CH {{\cal H}}
\def\CC {{\cal C}}
\def\CB {{\cal B}}
\def\CS {{\cal S}}
\def\CA{{\cal A}}
\def\CK{{\cal K}}
\def\CQ{{\cal Q}}

\def\p{\partial}
\def\pb{{\bar \p}}

\def\vt#1#2#3{ {\vartheta[{#1 \atop  #2}](#3\vert \tau)} }

\def\jb{{\bar j}}

\def\inbar{\,\vrule height1.5ex width.4pt depth0pt}

\def\half{{1 \over 2}}

\def\RR{{\text{R}}}

\newcommand{\cbrac}[1]{\left(#1\right)}
\newcommand{\sbrac}[1]{\left[#1\right]}
\newcommand{\ccbrac}[1]{\left\{#1\right\}}

\newcommand{\tr}{\text{Tr}}

\newcommand{\susy}[1]{$\mathcal{N}=#1$}
\newcommand{\zbar}{\bar{z}}
\newcommand{\PB}[1]{\left[#1\right]_{\text{PB}}}
\newcommand{\ta}{\widetilde{\alpha}}
\newcommand{\tbe}{\widetilde{\beta}}

\newcommand{\ga}{\acute{\alpha}}
\newcommand{\gb}{\acute{\beta}}

\newcommand{\bogmoduli}[1]{\mathcal{M}_{\text{B}}\left(G,#1\right)}

\newcommand{\e}[1]{\text{e}^{#1}}

\newcommand{\SL}[1]{SL\cbrac{#1}}

\newcommand{\su}[1]{\mathfrak{su}\cbrac{#1}}

\newcommand{\real}{\mathbb{R}}
\newcommand{\complex}{\mathbb{C}}
\newcommand{\integer}{\mathbb{Z}}

\newcommand{\Bc}{\mathcal{B}_{\text{c}}}
\newcommand{\Bop}{\mathcal{B}_{\text{Op}}}
\newcommand{\Bopp}{\mathcal{B}'_{\text{Op}}}
\newcommand{\Boper}{\mathcal{B}_{\text{Op}}}
\nc{\hTheta}{\hat{\Theta}}
\nc{\vp}{\varphi}
\nc{\tg}{\widetilde{g}}

\let\OLDthebibliography\thebibliography
\renewcommand\thebibliography[1]{
	\OLDthebibliography{#1}
	\setlength{\parskip}{5pt}
	\setlength{\itemsep}{0pt plus 0.3ex}
}
\usepackage{titlesec}
\titleformat*{\section}{\bfseries\large}
\begin{document}
\addtolength{\baselineskip}{1.5mm}

\thispagestyle{empty}
\vbox{}
\vspace{3.0cm}

\begin{center}
	\centerline{\LARGE{4d Chern-Simons Theory as a 3d Toda Theory,}}
	\bigskip
	\centerline{\LARGE{and a 3d-2d Correspondence}} 
	
\vspace{2.5cm}


		
	{Meer~Ashwinkumar\footnote{E-mail: meerashwinkumar@u.nus.edu}, Kee-Seng~Png\footnote{E-mail: pngkeeseng@u.nus.edu}, Meng-Chwan~Tan\footnote{E-mail: mctan@nus.edu.sg}}
	\\[2mm]
	{\it Department of Physics\\
		National University of Singapore \\
		2 Science Drive 3, Singapore 117551} \\[1mm] 
\end{center}

\vspace{2.0cm}

\centerline{\bf Abstract}\smallskip \noindent

We show that the four-dimensional Chern-Simons theory studied by Costello, Witten and Yamazaki, is,
 with Nahm pole-type boundary conditions, dual to a boundary theory 
 that is a 
three-dimensional analogue of Toda theory with a  novel 3d W-algebra symmetry. 
By embedding four-dimensional Chern-Simons theory in a partial twist of the five-dimensional maximally supersymmetric Yang-Mills theory on
a manifold with corners, we
 argue 
 that this three-dimensional Toda theory is dual to a
 two-dimensional topological sigma model with A-branes on the moduli space of solutions to the Bogomolny 
 equations.
 This furnishes a novel 3d-2d correspondence, which, among other mathematical implications, also reveals that modules of the 3d W-algebra are modules for the quantized algebra of certain holomorphic functions on  the Bogomolny moduli space. 


\newpage

\renewcommand{\thefootnote}{\arabic{footnote}}
\setcounter{footnote}{0}

\tableofcontents
\section{Introduction, Summary and Conventions}

\bigskip\noindent\textit{Introduction}
\vspace*{0.5em}


In 2010, 
the authors in \cite{alday2010liouville} observed a correspondence between two-dimensional Liouville theory and four-dimensional $\mathcal{N}=2$ super Yang-Mills (SYM) with gauge group $SU(2)$. This was termed the AGT correspondence.
This correspondence was further investigated for higher-rank gauge groups, where the correspondence is between two-dimensional Toda theory and four-dimensional  $\mathcal{N}=2$ SYM with gauge group $SU(N)$ \cite{wyllard20091}. Among other approaches, it is known that the AGT correspondence can be understood in terms of the 
characterization of the space of W-algebra conformal blocks as an irreducible module for quantized algebras of 
certain holomorphic functions on Hitchin's moduli space \cite{nekrasov2010omega}.
This implies a 
relationship between  Hitchin's moduli space and W-algebra representations.


In this work, we shall present an analogous relationship involving the moduli space of solutions to the Bogomolny equations (which is known to reduce to Hitchin's equations upon dimensional reduction), and a novel W-algebra arising from a three-dimensional analogue of Toda theory. The latter, as we shall show, arises as the boundary dual of the four-dimensional Chern-Simons theory studied by Costello, Witten and Yamazaki \cite{costello2017gaugeI,costello2018gaugeII} on $I\times S^1\times \Sigma$, with Nahm pole-type boundary conditions at the ends of the interval, $I$.

To show the aforementioned relationship, we shall first embed four-dimensional Chern-Simons theory in partially-twisted five-dimensional maximally supersymmetric Yang-Mills theory (MSYM) on $I\times\R_+\times S^1 \times \Sigma$, following \cite{ashwinkumar2019unifying} (the partial twist being analogous to the geometric Langlands twist of four-dimensional $\mathcal{N}=4$ SYM). Using the topological-holomorphic nature of the partially twisted theory, we scale up $I\times\R_+$ to arrive at a two-dimensional A-twisted supersymmetric sigma model on the moduli space of solutions to the Bogomolny equations defined on $ S^1\times \Sigma$. This implies a novel 3d-2d correspondence, that is, between the sigma model and the aforementioned three-dimensional Toda theory on $ S^1\times \Sigma$.

Let us now give a brief plan and summary of the paper.

\bigskip\noindent\textit{A Brief Plan and Summary of the Paper}
\vspace*{0.5em}

In \S\ref{sec:2}, we outline some basic properties of 5d MSYM on a manifold of the form $\CM=I\times\R_+\times S^1 \times {\Sigma}$, partially twisted along $I\times\R_+\times S^1$. We choose a topological twist analogous to the geometric Langlands twist in 4d \susy{4} super Yang-Mills. The action of the 5d ``GL''-twisted MSYM takes the form
\begin{equation*}
\boxed{S=\{\CQ,\til{V}\}+\frac{\til{\Psi}}{4\pi}\int_{I\times\R_+\times S^1\times \Sigma}\ dz\wedge\tr\cbrac{\cF \wedge \cF}}
\end{equation*}
In what follows, we shall take the gauge group to be $G=SU(N)$.

At the origin of $\R_+$, we shall choose NS5-type boundary conditions to maintain supersymmetry. At infinity along $\R_+$, the boundary conditions are taken to be $\CQ$-invariant configurations that are independent of the coordinate parameterizing $\R_+$.

At the two ends of the interval, we choose Nahm pole-type boundary conditions of the form
\begin{equation*}
\boxed{\begin{aligned}
&\CA\to {id\sigma\over \sigma}H+{dx^+\over \sigma}T_+\\
&\CA\to {-id\sigma\over \sigma-\pi}H+{dx^-\over \sigma-\pi}T_-
\end{aligned}}
\end{equation*}
where $x^\pm$ are certain light-cone-like coordinates on $S^1\times \Sigma$, and where $\sigma$ parameterizes the interval $I=[0,\pi]$. Here, $H$ and $T_\pm$ are the images of Cartan and ladder operators of an $\su{2}$ subalgebra of $\su{N}$, respectively. 

Furthermore, the corresponding path integral localizes to the path integral of the 4d Chern-Simons theory with gauge group $SL(N,\C)$, 
on $I\times S^1\times {\Sigma}$, which has the action
\begin{equation*}
\boxed{S_{\text{4d CS}}[\CA]={1\over2\pi\hbar}\int_{I\times S^1\times {\Sigma}}dz\wedge \Tr\cbrac{\CA\wedge d\CA+{2\over 3}\CA\wedge\CA\wedge\CA}}
\end{equation*}
where $z$ is a complex coordinate on $\Sigma$.

In \S\ref{sec:3}, we derive a 3d analogue of Toda theory from our partially-twisted 5d MSYM. Having localized to 4d Chern-Simons theory on $I\times S^1\times {\Sigma}$, 
the Nahm pole-type boundary conditions coincide with boundary conditions relating the 4d CS theory to a constrained 3d WZW model on each boundary. These constrained 3d WZW models are then shown to lead to the 3d Toda theory with action
\begin{equation*}
\boxed{\begin{aligned}
S_{\text{3d Toda}}[\phi]={1\over 2\pi\hbar}\int_{ S^1\times{\Sigma}}dz\wedge dx^+\wedge dx^-\,\cbrac{C_{ij}\p_+\phi^i\p_-\phi^j-4\sum_{i}\mu_i\nu_i\e{C_{ij}\phi_j}}
\end{aligned}}
\end{equation*}

Via an equivalent, gauged version of the 3d WZW model, we also derive a 3d analogue of the Virasoro algebra from 3d Liouville theory, i.e., the 3d Toda theory arising from the 4d CS theory with $\SL{2,\C}$ gauge group. 
This is of the form
\begin{equation*}
    \boxed{\begin{aligned}
    \sbrac{\Theta(z',\xi'),\Theta(z,\xi)}=
    &\left(\vphantom{\half}\p_\xi\Theta(z,\xi)\delta(\xi-\xi')+2\Theta(z,\xi)\delta'(\xi-\xi')\right.\\
    &\quad\qquad\left.+\cbrac{k-{\eta^2\over 2}}\delta'''(\xi-\xi')
    +k\delta'(\xi-\xi')
    \right)\delta(z-z')
    \end{aligned}}
\end{equation*}
where $\xi=x^+$ or $x^-$. This derivation can be generalized to $SL(N,\complex)$ gauge group in a straightforward manner, whereby we would obtain a 3d analogue of W-algebras.

In \S\ref{sec:4}, we derive a topological sigma model with A-branes from the partially twisted 5d MSYM. By exploiting the topological-holomorphic properties of the 5d ``GL''-twisted theory, we scale up $I\times\R_+$.  
The bosonic part of the 5d action then becomes the action of a sigma model of maps $X:I\times\R_+\to\bogmoduli{ S^1\times \Sigma}$ which takes the form 
\begin{equation*}
    \boxed{
    {S'}^{(A,\varphi)}_{\text{5d GL}}=
    \int_{I\times\R_+}d^2x\,G^{\text{B}}_{IJ}\p^{\ta}X^I\p_{\ta}X^J}
\end{equation*}
where $\bogmoduli{ S^1\times \Sigma}$ is the moduli space of solutions to the Bogomolny equations on $ S^1\times \Sigma$ which take the form
\begin{equation*}
    \boxed{\begin{aligned}
    &F_{\tau\zbar}-iD_{\zbar}\varphi_\tau=0\\
    &D_{\tau}\varphi^{\tau}-2iF_{z\zbar}=0
    \end{aligned}}
\end{equation*}
where $\tau$ parameterizes $S^1$. Furthermore, a term in the non-$\CQ$-exact sector of the 5d action reduces to an integral over a pullback to $I\times \R_+$ of a symplectic form on $\bogmoduli{ S^1\times \Sigma}$, i.e., $\omega = \til{\Psi}\omega_K^B$, with
\begin{equation*}
    \boxed{\omega^{\text{B}}_{K} = \frac{1}{2\pi} \int_{ S^1\times \Sigma} d^3x~ \textrm{Tr} (\delta \phi_\tau\wedge \delta A_4 - \delta A_5\wedge \delta A_\tau)}
\end{equation*}
where $A_4=A_z+A_{\zbar}$ and $A_5=i(A_z-A_{\zbar})$. 
This implies that the sigma model we derived is an A-model depending on the corresponding symplectic structure. We then compute the space of physical states of the A-model to be the space of certain holomorphic sections of bundles defined on Lagrangian branes denoted $\Bop$ and $\Bopp$, i.e., $H^0(\Bop,K^{1/2}_{\Bop})\otimes H^0(\Bopp,K^{1/2}_{\Bopp})$. These physical states are further identified with modules for the quantized algebra of certain holomorphic functions on $\bogmoduli{ S^1\times \Sigma}$.

\begin{figure}[t!]
	\centering
	\tikzset{every picture/.style={line width=0.75pt}} 

\begin{tikzpicture}[x=0.75pt,y=0.75pt,yscale=-1,xscale=1]

\draw   (206.32,258.11) .. controls (198.08,248.14) and (189.8,242.06) .. (181.51,239.87) .. controls (189.82,238.17) and (198.18,232.59) .. (206.55,223.12) ;
\draw    (205,235) -- (289,235) ;
\draw    (205,246) -- (289,246) ;
\draw    (229.51,72.76) -- (152.39,100.09) ;
\draw [shift={(150.51,100.76)}, rotate = 340.48] [color={rgb, 255:red, 0; green, 0; blue, 0 }  ][line width=0.75]    (10.93,-3.29) .. controls (6.95,-1.4) and (3.31,-0.3) .. (0,0) .. controls (3.31,0.3) and (6.95,1.4) .. (10.93,3.29)   ;
\draw    (144,171) -- (144,203.2) ;
\draw [shift={(144,205.2)}, rotate = 269.99] [color={rgb, 255:red, 0; green, 0; blue, 0 }  ][line width=0.75]    (10.93,-3.29) .. controls (6.95,-1.4) and (3.31,-0.3) .. (0,0) .. controls (3.31,0.3) and (6.95,1.4) .. (10.93,3.29)   ;
\draw    (318.68,73.21) -- (411.37,180.69) ;
\draw [shift={(412.68,182.21)}, rotate = 229.23] [color={rgb, 255:red, 0; green, 0; blue, 0 }  ][line width=0.75]    (10.93,-3.29) .. controls (6.95,-1.4) and (3.31,-0.3) .. (0,0) .. controls (3.31,0.3) and (6.95,1.4) .. (10.93,3.29)   ;
\draw   (288.51,257.09) .. controls (296.69,247.61) and (304.68,242.01) .. (312.45,240.3) .. controls (304.88,238.13) and (297.51,232.07) .. (290.35,222.14) ;

\draw    (214.95,3.66) -- (338.95,3.66) -- (338.95,72.66) -- (214.95,72.66) -- cycle  ;
\draw (276.95,38.16) node   [align=center] {
5d MSYM\\on\\$\displaystyle I  \times \mathbb{R}_+ \times S^1  \times \Sigma$
};
\draw    (102.95,102.66) -- (184.95,102.66) -- (184.95,171.66) -- (102.95,171.66) -- cycle  ;
\draw (143.95,137.16) node   [align=center] {
4d CS\\on\\$\displaystyle I\times  S^1 \times \Sigma$
};
\draw    (113.45,206.66) -- (174.45,206.66) -- (174.45,275.66) -- (113.45,275.66) -- cycle  ;
\draw (143.95,241.16) node   [align=center] {
3d Toda\\on\\$\displaystyle  S^1 \times \Sigma$
};
\draw    (320.45,185.66) -- (513.45,185.66) -- (513.45,282.66) -- (320.45,282.66) -- cycle  ;
\draw (416.95,234.16) node   [align=center] {
2d A-model\\on\\$\displaystyle I\times \mathbb{R}_+$\\with target $\bogmoduli{ S^1\times \Sigma}$
};
\draw (195,85.2) node [anchor=north west][inner sep=0.75pt]    {$\partial _{\mathbb{R}_+}$};
\draw (356,96) node [anchor=north west][inner sep=0.75pt]    {Scaling up $ I\times\R_+$};
\draw (146,176) node [anchor=north west][inner sep=0.75pt]    {${I}$};
\end{tikzpicture}
	\caption{Outline of the steps in this paper. 
		Starting from ``GL"-twisted 5d MSYM, we are able to establish a novel correspondence between the 3d Toda theory and a 2d A-twisted sigma model governing maps $I\times \R_+\to\bogmoduli{ S^1\times \Sigma}$.}
	\label{fig:5dsum}
\end{figure}
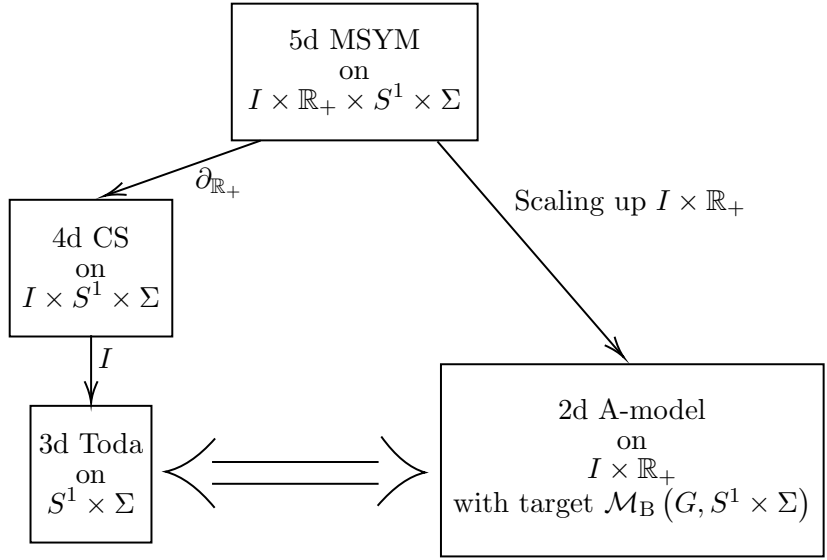

In \S\ref{sec:5}, via the topological-holomorphic property of the parent theory, i.e. partially twisted MSYM on $\CM=I\times\R_+\times S^1\times {\Sigma}$, we identify 3d Toda states with physical states of the 2d A-model with target space $\bogmoduli{S^1\times \Sigma}$. 
Hence, we establish a \textit{novel} correspondence between the 3d Toda theory and the 2d A-twisted sigma model with target space $\bogmoduli{S^1\times \Sigma}$, i.e., we have
\begin{equation*}
\boxed{\begin{aligned}
	&\text{3d Toda theory}\\ 
	&\quad\qquad\text{on}\\
	&\qquad S^1\times \Sigma
	\end{aligned}\qquad\Huge{\text{$\Leftrightarrow$}}\normalsize\text{}\qquad
	\begin{aligned}
	&\text{Topological sigma model with A-branes}\\ 
	&\quad\qquad\qquad\qquad\text{on}\\
	&\qquad\qquad\qquad I\times \R_+\\
	&\qquad\text{with target }\bogmoduli{ S^1\times \Sigma }
	\end{aligned}}
\end{equation*}
Mathematically, this implies that 
\begin{equation*}
\boxed{\begin{aligned}
	&\text{Modules of 3d W-algebras}\\ 
	&\qquad\quad\text{defined on}\\
	&\qquad\qquad S^1\times \Sigma
	\end{aligned}\quad\Huge{\text{$\Leftrightarrow$}}\normalsize\text{}\quad
	\begin{aligned}
	&\text{$H^0(\Bop,K^{1/2}_{\Bop})\otimes H^0(\Bopp,K^{1/2}_{\Bopp})$}\\ 
	\end{aligned}}
\end{equation*}
In other words, 
\begin{equation*}
    \boxed{\begin{aligned}
    &\quad\text{Modules of 3d W-algebras defined on $ S^1\times \Sigma$ are modules for the quantized algebra of}\quad\\
    &\quad\text{certain holomorphic functions on $\bogmoduli{ S^1\times \Sigma }$.}
    \end{aligned}}
\end{equation*}
\nn
The steps that we just outlined are summarized in Figure \ref{fig:5dsum}.

\newpage
\bigskip\noindent\textit{Labeling Conventions}
\vspace*{0.5em}

The labeling conventions for the indices used in this paper are as follows:

\renewcommand{\arraystretch}{1.5}
\begin{center}
	\begin{tabular}[p]{c|c}
Number of dimensions&Label\\ \hline
5d (on $I\times\R_+\times S^1\times {\Sigma}$)&$M,N,\cdots=1,2,3,4,5$\\\hline
3d (on $I\times\R_+\times S^1$) &$\alpha,\beta,\cdots=1,2,3$\\\hline
3d (on $ S^1\times{\Sigma}$)&$p,q,\cdots=3,4,5$\\\hline
2d (on ${I\times\R_+}$)&$\ta,\tbe,\cdots=1,2$\\\hline
2d (on ${I\times S^1}$)&$\ga,\gb,\cdots=1,3$
\end{tabular}
\end{center}

\bigskip\noindent\textit{Acknowledgments}
\vspace*{0.5em}

We would like to thank Sam Van Leuven, Gerben Oling, and Alessandro Tanzini for helpful correspondences. The results in this paper were presented at String Math 2020, and we would like to thank the audience, in particular, Nathan Seiberg, Cumrun Vafa, and Edward Witten for interesting comments and questions.
This work is supported by the MOE Tier 2 grant R-144-000-396-112.

\section{The Starting Point: Partially Twisted 5d MSYM\label{sec:2}}

\subsection{``GL"-twisted 5d MSYM\label{subsec:top twist}}


The classical action of 5d maximally supersymmetric Yang-Mills (MSYM) theory is of the form \cite{geyer2003higher,cordova2017five} 
\begin{equation}
\label{action:5d MSYM}
\begin{aligned}
S=&-{1\over {g_{\text{5d}}}^2}\int_{\CM} d^5x\,\tr\left(\quar F_{MN}F^{MN}
+\half D_M\varphi_{\hat{M}}D^M\varphi^{\hat{M}}
+\quar[\varphi_{\hat{M}},\varphi_{\hat{N}}][\varphi^{\hat{M}},\varphi^{\hat{N}}]\right.\\
&\qquad\qquad\qquad\qquad\left.
+i\rho^{\bar{M}\hat{\bar{M}}}{(\Gamma^M)_{\bar{M}}}^{\bar{N}}D_M\rho_{\bar{N}\hat{\bar{M}}}
+\rho^{\bar{M}\hat{\bar{M}}}{(\Gamma^{\hat{M}})_{\hat{\bar{M}}}}^{\hat{\bar{N}}}[\varphi_{\hat{M}},\rho_{\bar{M}\hat{\bar{N}}}]\right),
\end{aligned}
\end{equation}
where $M,N=1,\cdots,5$. Barred and hatted versions, i.e., $\bar{M}$ and $\hat{M}$, take the meaning of spinor and R-symmetry indices, respectively. This action is invariant under the supersymmetry transformations
\begin{subequations}
	\begin{eqnarray}
	&\delta A_M=i\zeta^{\bar{M}\hat{\bar{M}}}{\cbrac{\Gamma_M}_{\bar{M}}}^{\bar{N}}\rho_{\bar{N}\hat{\bar{M}}}\\
	&\delta\varphi^{\hat{M}}=\zeta^{\bar{M}\hat{\bar{M}}}{\cbrac{\Gamma^{\hat{M}}}_{\hat{\bar{M}}}}^{\hat{\bar{N}}}\rho_{\bar{M}\hat{\bar{N}}}\\
	&\begin{aligned}
	\delta\rho_{\bar{M}\hat{\bar{M}}}&=-i{\cbrac{\Gamma^M}_{\bar{M}}}^{\bar{N}}D_M\varphi^{\hat{M}}{\cbrac{\Gamma_{\hat{M}}}_{\hat{\bar{M}}}}^{\hat{\bar{N}}}\zeta_{\bar{N}\hat{\bar{N}}}
	-\half{\cbrac{\Gamma_{\hat{M}}}_{\hat{\bar{M}}}}^{\hat{\bar{N}}}{\cbrac{\Gamma_{\hat{N}}}_{\hat{\bar{N}}\hat{\bar{L}}}}[\varphi^{\hat{M}},\varphi^{\hat{N}}]{\zeta_{\bar{M}}}^{\bar{L}}\\
	&\qquad+\half F^{MN}{\cbrac{\Gamma_{MN}}_{\bar{M}}}^{\bar{N}}\zeta_{\bar{N}\hat{\bar{M}}}.
	\end{aligned}
	\end{eqnarray}
\end{subequations}


We shall take the underlying five-manifold to be 
of the form $\CM=V\times {\Sigma}$, where submanifolds $V$ and $\Sigma$ correspond to the $\ccbrac{x^1,x^2,x^3}$ and $\ccbrac{x^4,x^5}$ directions, respectively. In what follows, we shall also take the gauge group to be $G=SU(N)$. To be consistent with the conventions used in \cite{ashwinkumar2019unifying,kapustin2006electric}, we shall also take the Lie algebra $\mathfrak{g}$ to be anti-hermitian, i.e., $U^\dagger=-U$ where $U\in\mathfrak{g}$.

We would like to carry out a partial topological twist on the submanifold $V$. For the purpose of deriving the results seen in this paper, we shall choose the twist analogous to the geometric Langlands (GL) twist \cite{kapustin2006electric}. The symmetry group of 5d MSYM is $SO_{\CM}(5)\times SO_\RR(5)$, where the subscripts denote rotation and R-symmetry groups, respectively.


Prior to carrying out the topological twist, we decompose the rotation group as
\begin{equation}
\label{decom:L}
SO_{\CM}(5)\to SO_V(3)\times SO_\Sigma(2),
\end{equation}
and the R-symmetry group as
\begin{equation}
\label{decom:R}
SO_\RR(5)\to SO_\RR(3)\times SO_\RR(2),
\end{equation}
where $SO_\RR(3)\subset SO_\RR(5)$ rotates the scalar fields $\ccbrac{\varphi_{\hat{1}},\varphi_{\hat{2}},\varphi_{\hat{3}}}$.

Carrying out the topological twist described earlier amounts to taking the diagonal subgroup of $SO_V(3)$ and $SO_\RR(3)$, i.e., take 
\begin{equation}
\label{twist}
SO_V(3)'\subset SO_V(3)\times SO_\RR(3), 
\end{equation}so that the total symmetry group is now $SO_V(3)'\times SO_\Sigma(2)\times SO_\RR(2)$. The twisting of fermions which transform as $(\mathbf{2},\mathbf{2})$ under $SO_V(3)\times SO_\RR(3)$ means that they now transform as $\mathbf{1}\oplus\mathbf{3}$ under $SO_V(3)'$. Moreover, the scalar fields now transform as the components $\ccbrac{\varphi_{1},\varphi_{2},\varphi_{3}}$ of a one-form on $V$.

As shown in \cite{ashwinkumar2019unifying}, for an appropriate choice of supercharge, $\CQ$, that is scalar along $V$, this theory is the 5d analogue of the GL-twisted 4d \susy{4} SYM studied in \cite{kapustin2006electric}.\footnote{Such a partial topological twist has also been discussed conceptually by Elliot and Pestun \cite{elliott2019multiplicative}.} The  supercharge can be expressed as
\begin{equation}
\CQ=u \CQ_L+v\CQ_R,
\end{equation}
where $u,v\in\complex$, and the labels $L$ and $R$ are reminiscent of the corresponding left- and right- handed supersymmetries in 4d GL-twisted theory.  
By multiplying in a factor of $1/u$, supersymmetry variations can be rescaled so that the supercharges depend only on the ratio $t=v/u$, i.e., take
\begin{equation}
\CQ_t=\CQ_L+t\CQ_R,
\end{equation}
where the ratio $t\in\C\mathbb{P}^1$ is the analogue of the \textit{twisting parameter} in the 4d GL-twisted theory \cite{kapustin2006electric}.
In fact, taking $\Sigma=\R\times S^1$ or $T^2$, whereby the $x^5$ direction is $S^1$, we can dimensionally reduce along the circle to obtain precisely the transformations used by Kapustin and Witten in \cite{kapustin2006electric} via $A_5\to\varphi_4$, and similar identifications for the fermionic fields \cite{ashwinkumar2019unifying}.
Moreover, for $t\neq \pm i$, one finds that the $\CQ$-transformations only depend on $\Sigma$ via its complex structure.
Hence, such a theory is topological on the submanifold $V$, but has holomorphic dependence on the remaining two directions along ${\Sigma}$.

In what follows, we shall be interested in $V=I\times\R_+\times S^1$ and $\Sigma=\C\mathbb{P}^1$, $\C^\times$ or $\C/(\integer+\tau\integer)$. Such an $\CM$ is a manifold with two corners.
We shall also pick $t=-1$ for our purposes.
The action of the 5d ``GL''-twisted theory that we study can be written as
\begin{equation}\label{action00}
\boxed{S=\{\CQ,\til{V}\}+{\til{\Psi}\over 4\pi}\int_{I\times\R_+\times S^1\times \Sigma}\ dz\wedge\tr\cbrac{\cF \wedge \cF}}
\end{equation}
where $\widetilde{\Psi}$ is a real parameter, since we have selected $t=-1$. 
The curvature $\CF$ is defined in terms of the complex connection involving $\CA_{\alpha}=A_{\alpha}+i\varphi_{\alpha}$ and $\CA_{\zbar}=A_{\zbar}=\half(A_4+iA_5)$, where $z=x^4+ix^5$ and $\zbar=x^4-ix^5$ are local complex coordinates on $\Sigma$, and where $\alpha=1,2,3$. 
To arrive at \eqref{action00}, one partially twists the physical action \eqref{action:5d MSYM}, and finds that it can be expressed in terms of a $\CQ$-exact term, $\{\CQ,\til{V}\}$, and a term that is not $\CQ$-invariant due to the presence of the boundaries. Additional interaction terms ought to be added to arrive at \eqref{action00}, where the second term is $\CQ$-invariant, as long as $\CQ(\CA_{\alpha})=0$ and $\CQ(\CA_{\zbar})=0$ at the boundaries, and the corresponding connection obeys the Bianchi identity. The boundary conditions we choose satisfy these requirements, as we shall explain in the next subsection.




\subsection{Boundary Conditions}



Since our five-dimensional gauge theory is defined on a manifold with boundaries, we ought to specify boundary conditions at each of these boundaries.

Firstly, we shall choose an NS5-type boundary condition at the origin of $\R_+$.  This boundary condition requires that $t$ is real-valued, and includes the conditions that
\begin{subequations}
    \begin{eqnarray}
    &
    \CQ(\CA_{\ga})=0\\
    &
    \CQ(\CA_{\zbar})=0,
    \end{eqnarray}
\end{subequations}
where $\ga=1,3$, and $\CA_{\ga}$ and $\CA_{\zbar}$ obey Neumann boundary conditions. The fields $A_2$ and $\phi_2$ instead obey Dirichlet boundary conditions. This NS5-type boundary condition reduces, upon dimensional reduction along $x^5$ (for $\Sigma=\R\times S^1$ or $T^2$), to the ``topological'' NS5-type boundary condition considered in \cite{mikhaylov2018teichmuller}, where the choice of $t=-1$ implies that $\Psi$, the 4d analogue of $\til\Psi$, is real.

Secondly, at infinity along $\real_+$, the boundary conditions are taken to be $x^2$-independent $\CQ$-invariant configurations.





Thirdly, there are two boundaries at the endpoints of $I$, and we shall impose analogous boundary conditions for both of them. 
To define these boundary conditions we shall define the convenient coordinates
\begin{equation}
\label{eqn:lightcone}
   x^{\pm}=\tau\pm {i\over 2}\zbar,
\end{equation}
where $x^3=\tau$ parameterizes $S^1$.
The corresponding partial derivatives and complexified gauge fields are
\begin{align}
    &\p_\pm=\half\cbrac{\p_\tau\mp 2i\p_{\zbar}}\\
    &\CA_\pm=\half\cbrac{\CA_\tau\mp 2i\CA_{\zbar}},
\end{align}
respectively. 
The boundary conditions at the endpoints of the interval are Nahm pole-type boundary conditions, and can be defined as follows. As we approach one end of the interval, namely $\sigma\to 0$, where $x^1=\sigma\in I=[0,\pi]$, we impose boundary conditions that include
\begin{equation}
\label{con:nahm0}
\boxed{\CA\to {id\sigma\over \sigma}H+{dx^+\over \sigma}T_+}
\end{equation}
Here, for $\su{2}\subset \su{n}$, a homomorphism $\rho:\su{2}\to \su{N}$ has been chosen,
such that the image, $T_+$, of the raising operator of $\su{2}$ is a maximal length Jordan block, i.e.,
\begin{equation}
\label{eqn:image+}
T_+= 
-i
\begin{pmatrix}
0&\mu_1&0&\cdots &0\\
0&0&\mu_2&\cdots&0\\
\vdots &\vdots &\vdots&\ddots&\vdots\\
0&0&0&\cdots &\mu_{N-1}\\
0&0&0&\cdots&0
\end{pmatrix}.
\end{equation}
where the imaginary number $i$ accounts for the anti-hermiticity of 
 fields in the 5d ``GL''-twisted theory, and where $\mu_i$ are real and non-zero constants.
Imposing the boundary condition \eqref{con:nahm0} implies that
\begin{equation}
\label{con:A0}
\CA_{-}=0
\end{equation}
at $\si=0$.


On the other end of the interval, where $\sigma\to \pi$, we impose boundary conditions that include
\begin{equation}
\label{con:nahmpi}
\boxed{\CA\to {-id\sigma\over \sigma-\pi}H+{dx^-\over \sigma-\pi}T_-}
\end{equation}
Here, for $\su{2}\subset \su{n}$ we choose a homomorphism $\rho':\su{2}\to \su{N}$ such that the image, $T_-$, of the lowering operator is a maximal length Jordan block, i.e.,
\begin{equation}
\label{eqn:image-}
T_-=-
i
\begin{pmatrix}
0&0&\cdots &0&0\\
\nu_1&0&\cdots&0&0\\
0&\nu_2&\cdots &0&0\\
\vdots &\vdots&\ddots&\vdots&\vdots\\
0&0&\cdots&\nu_{N-1}&0
\end{pmatrix},
\end{equation}
where $\nu_i$ are real and non-zero constants. 
The boundary condition \eqref{con:nahmpi} implies that we set
\begin{equation}
\label{con:Api}
\CA_{+}=0
\end{equation}
at $\si=\pi$.

In order for these Nahm pole-type boundary conditions at the two ends of the interval to be well-defined, we also require that $\CA_{\si}$, $\CA_{\tau}$, and $\CA_{\zbar}$ are $\CQ$-invariant at these boundaries. It is crucial that all our boundary conditions are consistent at the corners of our five-manifold, and the fact that $\CA_{\alpha}$ and $\CA_{\zbar}$ are $\CQ$-invariant (or zero) at all the boundaries ensures this. 


It can be easily shown that upon dimensional reduction on $S^1\subset \Sigma$ (for $\Sigma=\R\times S^1$ or $T^2$), the boundary conditions \eqref{con:nahm0} and \eqref{con:nahmpi} reduce to
\begin{equation}
    \label{con:nahm0red}
    \CA\to {id\sigma\over \sigma}H+{du\over \sigma}T_+
\end{equation}
and 
\begin{equation}
    \label{con:nahmpired}
    \CA\to {-id\sigma\over \sigma-\pi}H+{d{\ov u}\over \sigma-\pi}T_-,
\end{equation}
respectively, where $u=\tau+ ix^4$ and $\ov{u}=\tau- ix^4$ are complex coordinates. 
These are known, from \cite{gaiotto2012knot}, to be complex Nahm pole boundary conditions for the 4d GL-twisted theory. Moreover, the resulting corner configurations at the intersection of the origin of $\R_+$ and the endpoints of the interval have been argued to be consistent in \cite{mikhaylov2018teichmuller}.


Having specified all our boundary conditions, we can observe that the $\CQ$-invariance of the second term in $\eqref{action00}$ follows from the fact that $\CA_{\alpha}$ and $\CA_{\zbar}$ are $\CQ$-invariant (or zero) at all the boundaries and the fact that the Bianchi identity is valid everywhere on the five-manifold. The latter is true since the Nahm pole-like configurations are complex flat connections, and therefore there are no singularities in their curvatures.

\subsection{Localization to 4d Chern-Simons Theory}


The path integral of the 5d partially-twisted theory localizes to solutions of $\CQ$-invariant configurations, whereby it reduces to a path integral over the non-$\CQ$-exact term in \eqref{action00}. Via Stokes' theorem, and the boundary conditions at each boundary, one finds that this 
term becomes a nontrivial $\CQ$-invariant boundary term at the origin of $\R_+$ (that takes the form of the 4d Chern-Simons action), together with a constant term at infinity along $\R_+$. The boundary conditions at the endpoints of $I$ set the 4d Chern-Simons actions there to zero.
As a result, up to normalization,
one arrives at
the
path integral of 4d Chern-Simons (CS) theory, which takes the form
\begin{equation}\label{4dcs}    \int_{\til{\Gamma}}D\CA\exp\cbrac{{i\over2\pi\hbar}\int_{I\times S^1\times {\Sigma}}dz\wedge \Tr\cbrac{\CA\wedge d\CA+{2\over 3}\CA\wedge\CA\wedge\CA}},
\end{equation}
where $\til{\Gamma}$ is the integration cycle defined by the $\CQ$-invariant localization equations, and ${i\over \hbar}={\til\Psi\over 2}$. This integration cycle is a Lefschetz thimble that ensures the convergence of the path integral. Here, the gauge connection is defined by $\CA=\CA_\sigma d\sigma+\CA_\tau d\tau+\CA_{\bar{z}}d\bar{z}$.

Hence, the ensuing theory on the boundary of $\real_+$ is the 4d Chern-Simons theory, which has the action
\begin{equation}
\label{action:4dcs}
\boxed{S_{\text{4d CS}}[\CA]={1\over2\pi\hbar}\int_{I\times S^1\times {\Sigma}}dz\wedge \Tr\cbrac{\CA\wedge d\CA+{2\over 3}\CA\wedge\CA\wedge\CA}}
\end{equation}

\section{3d Toda Theory from Partially Twisted 5d MSYM\label{sec:3}}

Consider the 4d CS theory on a four-manifold with a single boundary along the topological plane. If one of the complex gauge fields (namely $\CA_{\zbar}$) is set to vanish at the boundary, a 3d analogue of the chiral Wess-Zumino-Witten (WZW) model can be obtained \cite{ashwinkumar20203dwzw}.

In the 4d CS theory defined in \eqref{action:4dcs}, the underlying manifold has two boundaries at the ends of $I$, where we have Nahm pole-type boundary conditions that, among other things, also constrain some gauge fields to vanish, i.e., $\CA_-=0$ and $\CA_+=0$ at $\sigma=0$ and $\sigma=\pi$, respectively (see \eqref{con:A0} and \eqref{con:Api}). Thus, we expect that the 4d CS theory in our present analysis can be described by 3d WZW models at the boundaries, with further constraints coming from the Nahm pole-type boundary conditions.

In what follows, we shall first derive 3d WZW models defined at the boundaries of $I$, and then include the further constraints that originate from the Nahm pole-type boundary conditions to obtain a 3d analogue of analytically-continued Toda theory.

\subsection{Derivation of 3d WZW Model Description of Boundary Theory\label{sec:WZW}}

\bigskip\noindent\textit{3d WZW Model at $\sigma\to0$}
\vspace*{0.5em}

We shall first verify the locality and gauge invariance of the 4d CS theory defined in \eqref{action:4dcs}, near the boundary at $\sigma=0$. 
To show the locality of our 4d CS theory here, we first vary the action \eqref{action:4dcs} to give
\begin{equation}
\label{action:variation4dCS}
\delta S_{\text{4d CS}}={1\over 2\pi\hbar}\int_{I\times S^1\times{\Sigma}}dz\wedge\Tr\cbrac{\delta\CA\wedge\CF+d(\delta\CA\wedge\CA)}.
\end{equation}
Using Stokes' theorem, the second term of the variation \eqref{action:variation4dCS} is identified as a boundary term. Imposing the boundary condition $\CA_{-}=0$ (see \eqref{con:A0}), the boundary variation vanishes.

To show the gauge invariance of our 4d CS theory, we first extend the partial connection $\CA$ to a full connection, which allows us to rewrite \eqref{action:4dcs} as
\begin{equation}
\label{action:4d CSv2}
S_{\text{4d CS}}[\CA]=-{1\over2\pi\hbar}\int_{I\times S^1\times {\Sigma}}z\tr\cbrac{\CF\wedge\CF}+{1\over2\pi\hbar}\int_{ S^1\times{\Sigma}}z\tr\cbrac{\CA\wedge d\CA+{2\over 3}\CA\wedge\CA\wedge\CA}.
\end{equation}
Using the boundary condition $\CA_{-}=0$, and imposing the additional boundary condition $\CA_z=0$, the second term of \eqref{action:4d CSv2} vanishes (it should be noted that we are not taking $\CA_z$ to be the complex conjugate of $\CA_{\zbar}$). We then find that the first term is gauge-invariant under large gauge transformations of the form
\begin{equation}
\label{eqn:gaugetrans}
    \CA\to U\CA U^{-1}-dUU^{-1}.
\end{equation}
However, we ought to restrict $U$ such that the boundary conditions $\CA_-=0=\CA_z$ are preserved. We shall achieve this by taking $U$ to tend to the identity element of $G$ at $\sigma=0$.

Now, in order to derive the 3d WZW model at the boundary at $\sigma=0$, we exploit the fact that the 4d CS action \eqref{action:4dcs} admits an alternative form
\begin{equation}
\label{action:4d CSv3}
{1\over2\pi\hbar}\int_{I\times S^1\times {\Sigma}}dz\wedge d\sigma \wedge dx^- \wedge dx^+\,\Tr\cbrac{2\CA_- \CF_{+\sigma}-\CA_{\sigma}\p_-\CA_++\CA_+\p_-\CA_\sigma },
\end{equation}
as $\sigma\to0$, where it can be seen that $\CA_-$ is a Lagrange multiplier. 
In turn, the Lagrange multiplier is then integrated out to produce the condition $\CF_{+\sigma}=0$, which has solutions that can be expressed as
\begin{subequations}
	\label{solns:zbar=0}
	\begin{eqnarray}
	&\CA_{\sigma}=g^{-1}\partial_{\sigma} g\\
	&\CA_+=g^{-1}\p_+g,
	\end{eqnarray}
\end{subequations}
where $g$ is a map, 
$g:I\times S^1\times {\Sigma}\to SL{(N,\C)}$. 
Changing variables from $\CA_\sigma$ and $\CA_+$ to $g$, \eqref{action:4dcs} then becomes
\begin{equation}
\label{action:3dwzw}
\begin{aligned}
\lim_{\sigma\to 0}S_{\text{4d CS}}[\CA]&=S_{\text{3d WZW}}[g]\\
&={1\over2\pi\hbar}\int_{ S^1\times{\Sigma}}\, dz\wedge dx^+\wedge dx^-\, \Tr\cbrac{\p_+ g\, g^{-1}\p_-g\, g^{-1}}\\
&\quad-{1\over 6\pi\hbar}\int_{I\times S^1\times {\Sigma}}\, dz\wedge\Tr\cbrac{dg\, g^{-1}\wedge dg\, g^{-1}\wedge dg\, g^{-1}},
\end{aligned}
\end{equation}
which takes the form of a 3d analogue of the 2d chiral WZW model. 

Next, we note that a gauge transformation \eqref{eqn:gaugetrans} amounts to $g \to g U^{-1}$ in \eqref{solns:zbar=0}. As a result, we may change the value of $g$ in the bulk without changing its value at $\sigma= 0$. This means that the Wess-Zumino term in \eqref{action:3dwzw} does not depend on the choice of extension of the boundary value of $g$ over the bulk manifold. 
 Thus, we can divide out the volume of the gauge group to obtain a path integral of the form
\begin{equation}
    \int D\tg\,\e{iS[\tg]},
\end{equation}
where $\tg$ is now the map $\tg:S^1\times \Sigma\to SL(N,\C)$.

Varying the action \eqref{action:3dwzw}, we obtain
\begin{equation}
\label{var:3dwzw}
\delta S_{\text{3d WZW}}=-{1\over\pi\hbar}\int_{ S^1\times{\Sigma}}dz\wedge dx^+\wedge dx^-\,\tr\cbrac{\tg^{-1}\delta \tg\p_-(\tg^{-1}\p_+\tg)},
\end{equation}
which implies the equation of motion
\begin{equation}
\label{eom:BC1t}
\p_-(\tg^{-1}\p_+\tg)=0.
\end{equation}
This equation of motion is equivalent to 
\begin{equation}
\label{eom:BC1zbar}
\p_+(\p_-\tg\, \tg^{-1})=0.
\end{equation}
Hence, the solutions of \eqref{eom:BC1t} and \eqref{eom:BC1zbar} take the form
\begin{equation}
\label{eqn:WZWzeromodes}
\tg(z,x^+,x^-)=A(z,x^+)B(z,x^-).
\end{equation}
The equations of motion, \eqref{eom:BC1t} and \eqref{eom:BC1zbar}, are the current conservation equations for the symmetry of the action under the transformation
\begin{equation}
\tg(z,x^+,x^-)\to\widetilde{\Omega}(z,x^+)\tg\,\Omega(z,x^-),
\end{equation}
where the conserved currents take the form
\begin{subequations}
	\label{eqn:3dwzw a-currents(zbar)}
	\begin{eqnarray}
	\label{eqn:2dwzw a-currents(zbar)a}
	&J_{+}=\tg^{-1}\,\p_+ \tg\\
	\label{eqn:2dwzw a-currents(zbar)b}
	&J_{-}=\p_- \tg\, \tg^{-1}.
	\end{eqnarray}
\end{subequations}
Crucially, it is seen that we can identify $\CA_+$ with $J_+$ on the boundary at $\sigma=0$, and so, the rest of the boundary conditions in \eqref{con:nahm0} also constrain $J_+$.


\bigskip\noindent\textit{3d WZW Model at $\sigma\to \pi$}
\vspace*{0.5em}

We shall now verify the locality and gauge invariance of the 4d CS theory defined in \eqref{action:4dcs}, near the boundary at $\sigma=\pi$. 
To show the locality of our 4d CS theory here, we first vary the action \eqref{action:4dcs}. Then, we impose the boundary condition $\CA_+=0$ (see \eqref{con:Api}). In doing so, the boundary contribution to the variation \eqref{action:variation4dCS} vanishes.

To show the gauge invariance of our 4d CS theory, we first extend the partial connection $\CA$ to a full connection. Using the boundary condition $\CA_{+}=0$, and imposing the additional boundary condition $\CA_z=0$, the second term of \eqref{action:4d CSv2} vanishes. Like before, we ought to restrict $U$ such that the boundary conditions $\CA_+=0=\CA_z$ are preserved. We shall achieve this by taking $U$ to tend to the identity element of $G$ at $\sigma=\pi$.
 
Now, in order to derive the 3d WZW model at the boundary at $\sigma=\pi$, we make use of the fact that the 4d CS action \eqref{action:4dcs} admits an alternative form
\begin{equation}
{1\over2\pi\hbar}\int_{I\times S^1\times {\Sigma}}dz\wedge d\sigma \wedge dx^- \wedge dx^+\,\Tr\cbrac{2\CA_+ \CF_{-\sigma}+\CA_{\sigma}\p_+\CA_--\CA_-\p_+\CA_\sigma },
\end{equation}
as $\sigma\to\pi$. It can be seen that $\CA_+$ is a Lagrange multiplier, which is then integrated out to produce the condition $\CF_{\sigma -}=0$. We shall choose solutions to this condition that take the form
\begin{subequations}
	\begin{eqnarray}
	\label{soln:t=0}
	&\CA_{\sigma}=-\partial_{\sigma}g\,g^{-1}\\
	&\CA_-=-\p_-g\,g^{-1},
	\end{eqnarray}
\end{subequations}
where $g$ is a map, 
$g:I\times S^1\times {\Sigma}\to SL{(N,\C)}$. 

Changing variables from $\CA_\sigma$ and $\CA_-$ to $g$, and taking note of the orientation of $I$, \eqref{action:4dcs} then becomes
\begin{equation}
\label{action:3dwzw2}
\begin{aligned}
\lim_{\sigma\to \pi}S_{\text{4d CS}}[\CA]&=S_{\text{3d WZW}}'[g]\\
&={1\over2\pi\hbar}\int_{ S^1\times{\Sigma}}\, dz\wedge dx^+\wedge dx^-\, \Tr\cbrac{\p_+ g\, g^{-1}\p_-g\, g^{-1}}\\
&\quad+{1\over 6\pi\hbar}\int_{I\times S^1\times {\Sigma}}\, dz\wedge\Tr\cbrac{dg\, g^{-1}\wedge dg\, g^{-1}\wedge dg\, g^{-1}},
\end{aligned}
\end{equation}
which differs from \eqref{action:3dwzw} by a minus sign on the topological term. 

Next, we note that a gauge transformation \eqref{eqn:gaugetrans} amounts to $g \to Ug$ in \eqref{soln:t=0}. As a result, we may change the value of $g$ in the bulk without changing its value at $\sigma= \pi$. This means that the Wess-Zumino term in \eqref{action:3dwzw2} does not depend on the choice of extension of the boundary value of $g$ over the bulk manifold. 
Thus, we can divide out the volume of the gauge group to obtain a path integral of the form
\begin{equation}
    \int D\tg\,\e{iS[\tg]},
\end{equation}
where $\tg$ is now the map $\tg:S^1\times \Sigma\to SL(N,\C)$.

Varying the action \eqref{action:3dwzw2}, and taking note of the orientation of $I$, we obtain
\begin{equation}
\delta S_{\text{3d WZW}}'=-{1\over\pi\hbar}\int_{ S^1\times{\Sigma}}dz\wedge dx^+\wedge dx^-\,\tr\cbrac{\tg^{-1}\delta \tg\p_-(\tg^{-1}\p_+\tg)},
\end{equation}
which is the same as \eqref{var:3dwzw}. This means that identical equations of motion, \eqref{eom:BC1t} and \eqref{eom:BC1zbar}, are obtained, and the corresponding conserved currents are proportional to those in \eqref{eqn:3dwzw a-currents(zbar)}.

With a convenient overall factor, the conserved currents are
\begin{subequations}
	\label{eqn:3dwzwcurrents pi}
	\begin{eqnarray}
	\label{eqn:3dwzwcurrents pia}
	&J'_{+}=-\tg^{-1}\,\p_+ \tg\\
	\label{eqn:3dwzwcurrents pib}
	&J'_{-}=-\p_- \tg\, \tg^{-1},
	\end{eqnarray}
\end{subequations}
where the prime on $J'$ reminds the reader that these currents are localized to $\sigma\to\pi$. Since we can identify $\CA_{-}$ with $J'_{-}$ on the boundary at $\sigma=\pi$, the rest of the boundary conditions in \eqref{con:nahmpi} also constrain $J'_{-}$.

\subsection{Derivation of 3d Toda Theory from Constrained 3d WZW Models\label{sec:3dToda}}


\bigskip\noindent\textit{Current Constraints}
\vspace*{0.5em}

Note that $T_+$ in \eqref{eqn:image+} ($T_-$ in \eqref{eqn:image-}) can be expressed as a sum
of the positive (negative) simple roots with non-zero coefficients. In addition, a current, $J$, can be projected onto Cartan directions, as well as roots of the Lie algebra, i.e., we can write
\begin{equation}
J=\sum_{\alpha\in\Delta}(J^{-\alpha} {R}_{-\alpha}+J^{+\alpha} R_{+\alpha})+J^{0i} R_{0i},
\end{equation}
where $R_{0i}$ denotes the $i$-th Cartan generator for $i=1,\cdots,N-1$, while $\Delta$ denotes the root space. 
Since $\mu_i$ and $\nu_i$ are non-zero only for simple roots, $R_{\pm i}$, the Nahm pole-type boundary conditions \eqref{con:nahm0} and \eqref{con:nahmpi} can be shown to imply that
\begin{subequations}
	\label{eqn:current constraints(t)}
	\begin{eqnarray}
	\label{eqn:current constraints(t)a}
		&{J_{+}}^{+i}=\mu_i\\
		\label{eqn:current constraints(t)b}
		&{J_{+}}^{0i}=0
	\end{eqnarray}
\end{subequations}
and
\begin{subequations}
	\label{eqn:current constraints(zbar)}
	\begin{eqnarray}
	\label{eqn:current constraints(zbar)a}
	&{J'_{-}}^{-i}=\nu_i\\
	\label{eqn:current constraints(zbar)b}
	&{J'_{-}}^{0i}=0,
	\end{eqnarray}
\end{subequations}
respectively. Together, \eqref{eqn:current constraints(t)} and \eqref{eqn:current constraints(zbar)} are current constraints. In this sense, what we really derived earlier is a \textit{constrained} 3d WZW model.

\bigskip\noindent\textit{Gauss Decomposition}
\vspace*{0.5em}

Using the Gauss decomposition, the field $\tg$ can be expressed as 
\begin{equation}
\label{eqn:gauss decomposition}
\tg=\exp\cbrac{i\sum_{\alpha\in\Delta}X_\alpha R_\alpha^+}\exp\cbrac{i\phi_i R_i^0}\exp\cbrac{i\sum_{\alpha\in\Delta}Y_\alpha R^-_\alpha},
\end{equation}
where $X_\alpha$, $Y_\alpha$ and $\phi_i$ are scalar fields, and summations are carried out over repeated indices. 
Using \eqref{eqn:gauss decomposition}, $J_{+}=\tg^{-1}\,\p_+ \tg$ may then be rewritten as
\begin{equation}
\label{eqn:current gauss decomposed (t)}
\begin{aligned}
J_{+}=&\sum_{i}\cbrac{i\p_+X_i\e{-C_{ij}\phi_j}}R_i^+
+\cbrac{-\sum_{i,k}C_{ik}Y_iY_k\p_+X_k\e{-C_{kj}\phi_j}-i\sum_{i,j}C_{ij}Y_i\p_+\phi_j+i\sum_{i}\p_+Y_i}R_i^-\\
&+\sum_{i}\cbrac{i\p_+\phi_i+2iY_i\p_+X_i\e{-C_{ij}\phi_j}}R_i^0+\cdots,
\end{aligned}
\end{equation}
where $C_{ij}$ are elements of the Cartan matrix, $\exp\cbrac{C_{ij}\phi_j}=\exp\cbrac{\sum_{j} C_{ij}\phi_j}$, and where $\cdots$ are terms involving non-simple roots. For $SL(N,\C)$, the Cartan matrix is written as
\begin{equation}
\sbrac{C_{ij}}=
\begin{pmatrix}
2&-1&\cdots&0&0\\
-1&2&\cdots&0&0\\
\vdots&\vdots&\ddots&\vdots&\vdots\\
0&0&\cdots&2&-1\\
0&0&\cdots&-1&2
\end{pmatrix}.
\end{equation}
Comparing \eqref{eqn:current gauss decomposed (t)} with \eqref{eqn:current constraints(t)}, we obtain
\begin{subequations}
	\label{eqn:X}
	\begin{eqnarray}
	&\p_+X_i\e{-C_{ij}\phi_j}=-i\mu_i\\
	&\p_+\phi_i+2Y_i\p_+X_i\e{-C_{ij}\phi_j}=0.
	\end{eqnarray}
\end{subequations}
Likewise, $J'_{-}=-\p_- \tg\, \tg^{-1}$ may be rewritten as
\begin{equation}
\label{eqn:current gauss decomposed (zbar)}
\begin{aligned}
J'_{-}=&\cbrac{\sum_{i,k}C_{ik}X_iX_k\p_-Y_k\e{-C_{kj}\phi_j}+i\sum_{i,j}C_{ij}X_i\p_-\phi_j-i\sum_{i}\p_-X_i}R_i^+
+\sum_{i}\cbrac{-i\p_-Y_i\e{-C_{ij}\phi_j}}R_i^-\\
&\quad+\sum_{i}\cbrac{-i\p_-\phi_i-2i X_i\p_-Y_i\e{-C_{ij}\phi_j}}R_i^0+\cdots,
\end{aligned}
\end{equation}
where $\cdots$ are terms involving non-simple roots. Comparing \eqref{eqn:current gauss decomposed (zbar)} with \eqref{eqn:current constraints(zbar)}, we obtain
\begin{subequations}
	\label{eqn:Y}
	\begin{eqnarray}
	&\p_-Y_i\e{-C_{ij}\phi_j}=i\nu_i\\
	&\p_-\phi_i+2X_i\p_-Y_i\e{-C_{ij}\phi_j}=0.
	\end{eqnarray}
\end{subequations}
Combining \eqref{eqn:X} and \eqref{eqn:Y} then gives
\begin{equation}
\label{eom:3dtoda}
\p_+\p_-\phi_i+2\mu_i\nu_i\e{C_{ij}\phi_j}=0,
\end{equation}
which resemble the equations of motion for 2d Toda theory. The corresponding action that gives these constraints as equations of motion is
\begin{equation}
\label{action:3dtoda}
\boxed{\begin{aligned}
S_{\text{3d Toda}}[\phi]={1\over 2\pi\hbar}\int_{ S^1\times{\Sigma}}dz\wedge dx^+\wedge dx^-\,\cbrac{C_{ij}\p_+\phi^i\p_-\phi^j-4\sum_{i}\mu_i\nu_i\e{C_{ij}\phi_j}}
\end{aligned}}
\end{equation}
which takes the form of a 3d analogue of analytically-continued 2d Toda theory, and is expected to be an integrable field theory since \eqref{eom:3dtoda} resembles the  equations of motion of 2d Toda theory, which can be solved exactly. Hence, the edge modes of the 4d CS theory with Nahm-pole-type boundary conditions are 3d Toda fields. 

One may also obtain 3d Toda theory from 4d CS theory via an off-shell procedure. The two constrained 3d `chiral' WZW models can be identified with a single 3d WZW model due to the topological invariance along $I$, and the Gauss decomposition \eqref{eqn:gauss decomposition} can then be substituted into it. Subsequently implementing the constraints \eqref{eqn:X} and \eqref{eqn:Y} in the resulting action furnishes the 3d Toda action \eqref{action:3dtoda}.

It is expected that if one includes Wilson lines along $I$ or 't Hooft operators along $I\times \R_+$
 in the 5d ``GL"-twisted theory, one should obtain Wilson and 't Hooft lines along $I$ in the 4d Chern-Simons theory that results from localization. These in turn ought to correspond to local operators in the dual 3d Toda theory.

\bigskip\noindent\textit{Dimensional Reduction}
\vspace*{0.5em}

If we take $\Sigma=\R\times S^1$ or $T^2$, one can carry out a dimensional reduction on the circle in the $x^5$ direction. The dimensional reduction amounts to setting $\partial_{\bar{z}}\to\half\partial_4$ and $dz\wedge d\zbar=-2i dx^4\wedge dx^5$, which implies that $\p_+\to{\p}_u=\half(\p_\tau- i\p_4)$ and $\p_-\to{\p}_{\ov u}=\half(\p_\tau+ i\p_4)$. The 3d Toda action \eqref{action:3dtoda} then becomes
\begin{equation}
\label{action:2d toda}
S_{\text{2d Toda}}[\phi]={ i R\over 2\hbar}\int_{C}du\wedge d{\ov u}\,\cbrac{C_{ij}{\p}_{u}\phi^i{\p}_{\ov u}\phi^j-4\sum_{i}\mu_i\nu_i\e{C_{ij}\phi_j}},
\end{equation}
where we have denoted the remaining two directions by $C$, and $R$ is the radius of the shrunken circle. 
The action \eqref{action:2d toda} indeed takes the form of the usual analytically-continued 2d conformal Toda action, as expected.


\subsection{Three-dimensional W-algebras}

In this section, we shall explain how the 3d Toda theory gives rise to analogues of W-algebras one usually obtains from 2d Toda theory. 
To this end, we shall first derive the 3d analogue of the 2d improved energy-momentum tensor, which gives rise to the Virasoro algebra in 2d Toda theory.

\bigskip\noindent\textit{A 3d Analogue of Improved Energy-momentum Tensor}
\vspace*{0.5em}

Unlike 2d Toda theory, the 3d Toda action \eqref{action:3dtoda} is not invariant under coordinate transformations of the form $x^p\to f(x^p)$, if we assume that $f(x^p)$ are all non-trivial transformations. This is due to the volume element, which transforms as $d^3x\to(\p_+f)(\p_-f)(\p_z f)d^3x$, while the 3d Toda Lagrangian transforms as $\CL_{\text{3d Toda}}\to{1\over (\p_+f)(\p_-f)}\CL_{\text{3d Toda}}$. Instead, we shall take only $f(x^+)$ to be a non-trivial transformation,\footnote
{  In the following analysis, we shall take $x^+$ and $x^-$ to be independent coordinates for simplicity. In doing so, we extend the range of the coordinate $\tau$ such that it is complex-valued. The physical space of interest, however, is defined by $\bar{\tau}=\tau$.
}  i.e., take non-trivial transformations only along the $x^+$-direction.\footnote
{
	It should also be noted that the choice $f=f(x^-)$ is equally valid. In fact, this alternate choice will lead to another copy of the W-algebras that we will derive from the 3d Toda theory.
}

Let us now consider infinitesimal coordinate transformations $f(x^+)=x^++\varepsilon(x^+)$ where $\varepsilon(x^+)\ll 1$. Since the transformation law for scalars is
\begin{equation}
\phi_i(z,x^+,x^-)=\phi'_i(z,x'^+,x^-)=\phi'_i(z,x^++\varepsilon(x^+),x^-),
\end{equation}
one would naively look at the corresponding scalar field transformation
\begin{equation}
    \delta\phi_i(z,x^+,x^-)=\phi'_i(z,x'^+,x^-)-\phi_i(z,x^+,x^-)=-\varepsilon\p_+\phi_i(z,x^+,x^-).
\end{equation}
However, while the kinetic term in \eqref{action:3dtoda} remains invariant under this transformation, the potential term does not.

Instead, we shall consider the following modified transformation law for the 3d Toda fields
\begin{equation}
\begin{aligned}
\phi_i'(z,x'^+,x^-)&= \phi_i(z,x^+,x^-)+\cbrac{\sum_j C^{-1}_{ij}}\ln \p_+ f\\
&=\phi_i(z,x^+,x^-)+\gamma_i\ln \p_+ f,
\end{aligned}
\end{equation}
where we have written $\sum_j C^{-1}_{ij}=\gamma_i$. It can be shown that the 3d Toda action remains invariant under the corresponding infinitesimal variations
\begin{equation}
\label{eqn:phivar}
\phi_i(z,x^+,x^-)\to \phi_i(z,x^+,x^-)-\varepsilon\p_+\phi_i+\gamma_i\p_+\varepsilon,
\end{equation}
whereby the Lagrangian varies as a boundary term.

The infinitesimal symmetry variation of the action \eqref{action:3dtoda} is given explicitly by
\begin{equation}
\label{var:3dtodasym}
\begin{aligned}
\delta S_{\text{3d Toda}}&=\int_{ S^1\times \Sigma}dz\wedge dx^+\wedge dx^-\,\left[\p_+\cbrac{-{1\over2\pi\hbar}\cbrac{C_{ij}\varepsilon\p_+\phi^i\p_-\phi^j-2\sum_i\varepsilon\p_+\p_-\phi_i}}\right.\\
&\qquad\qquad\qquad\qquad\qquad\qquad\qquad\qquad\qquad\qquad
\left.-\p_-\cbrac{{1\over2\pi\hbar}C_{ij}\gamma^i\p_+\p_+\varepsilon\phi^j}\right].
\end{aligned}
\end{equation}
This is equivalent to the on-shell variation of the action, which is written as
\begin{equation}
\label{var:3dtodaos}
{\delta}' S_{\text{3d Toda}}=\int_{ S^1\times \Sigma}dz\wedge dx^+\wedge dx^-\,\p_p\cbrac{{\p\CL_{\text{3d Toda}}\over\p\cbrac{\p_p\phi_i}}\delta\phi_i}.
\end{equation}
We can then derive the Noether current that corresponds to the symmetry of the action under the variation described in \eqref{eqn:phivar}, by equating the integrands of \eqref{var:3dtodasym} and \eqref{var:3dtodaos}, to obtain
\begin{equation}
\label{eqn:improvedconservation}
\p_-\sbrac{\cbrac{{1\over\pi\hbar}C_{ij}\cbrac{\half\p_+\phi^i\p_+\phi^j
-\gamma^i\p_+\p_+\phi^j}}}=0.
\end{equation}
Hence, we obtain the Noether current of the form
\begin{equation}
\label{em:improved}
{\Theta}={1\over2\pi\hbar}C_{ij}\p_+\phi^i\p_+\phi^j-{1\over\pi\hbar}C_{ij}\gamma^i\p_+\p_+\phi^j,
\end{equation}
that obeys the current conservation equation $\p_-{\Theta}=0$, whereby it must be true that ${\Theta}={\Theta}(z,x^+)$. The expression \eqref{em:improved} is the 3d analogue of a component of the improved energy-momentum tensor in 2d Toda theory. 

Next, we wish to use the Noether current $\Theta(z,x^+)$ to compute a 3d analogue of a W-algebra. We shall do so via an equivalent theory, namely a gauged version of the 3d WZW model.

\bigskip\noindent\textit{Equivalence between Constrained and Gauged 3d WZW Models}
\vspace*{0.5em}

We shall show that the constrained 3d WZW models, derived in \S\ref{sec:WZW}, can also be described by a \textit{gauged} 3d WZW model which has the action
\begin{equation}
\label{action:GWZW}
\begin{aligned}
S_{\text{3d GWZW}}[\tg,\fA_+,\fA_{-}]
=&S_{\text{3d WZW}}[\tg]
+{1\over 2\pi\hbar}\int_{ S^1\times \Sigma} dz\wedge dx^+\wedge dx^-\, \tr\left(\fA_+(\p_-\tg)\tg^{-1}+(\tg^{-1}\p_+ \tg)\fA_{-}
\right.\\
&\left.\qquad\qquad\qquad\qquad\qquad\qquad\qquad\qquad\qquad\qquad
+\fA_+ \tg\fA_{-}\tg^{-1}
-\fA_{-}\mu+\fA_{+}\nu\right),
\end{aligned}
\end{equation}
where the auxiliary gauge connections are defined by $\fA_+=h^{-1}\p_+ h$, $\fA_{-}=\widehat{h}^{-1}\p_- \widehat{h}$ and $h,\widehat{h}\in H\subset SL(N,\complex)$, while $\mu=\sum_{i}\mu_i R^+_{i}$ and $\nu=\sum_{i}\nu_i R^-_{i}$. Recall that $R^\pm_i$ denote simple roots of $\mathfrak{sl}(N,\complex)$.


The action \eqref{action:GWZW} can be shown to be invariant under the gauge transformations 
\begin{subequations}
	\begin{eqnarray}
	&\tg\to\alpha \tg\beta^{-1},&\alpha\in N^-, \beta\in N^+\\
	&h\to h\alpha^{-1}&\\
	&\widehat{h}\to \widehat{h}\beta^{-1},&
	\end{eqnarray}
\end{subequations}
where $N^\pm$ are the subspaces of $SL(N,\complex)$ that are generated by ladder operators associated with $\pm$ roots. 
Next, we take a variation of \eqref{action:GWZW} with respect to the fields $\fA_{-}$, $\fA_{+}$ and $\tg$ which leads to the equations of motion
\begin{subequations}
	\label{eom:GWZW}
	\begin{eqnarray}
	&\tg^{-1}\p_+ \tg+\tg^{-1}\fA_+ \tg=\mu\\
	&-\p_-\tg \tg^{-1}-\tg\fA_{-}\tg^{-1}=\nu\\
	&\begin{aligned}
	-2\p_-\cbrac{\tg^{-1}\p_+ \tg}
	&-\tg^{-1}\p_-\fA_{+}\tg
	+\tg^{-1}\p_-\tg\tg^{-1}\fA_{+}\tg
	-\tg^{-1}\fA_+ \p_-\tg
	-\tg^{-1}\p_+ \tg\fA_{-}\\
	&\qquad\qquad+\fA_{-}\tg^{-1}\p_+ \tg
	-\p_+ \fA_{-}
	+\fA_{-}\tg^{-1}\fA_+ \tg
	-\tg^{-1}\fA_+ \tg\fA_{-}
	=0,
	\end{aligned}
	\end{eqnarray}
\end{subequations}
respectively.

Finally, picking the partial gauge-fixing condition as $\fA_{+}=0=\fA_{-}$,\footnote
{
	Note that the gauge-fixing condition $\fA_{+}=0=\fA_{-}$ does not fix the gauge completely. There is a residual gauge symmetry $g\to \alpha g\beta^{-1}$, where $\alpha=\alpha(x^+)$ and $\beta=\beta(x^-)$ now depend only on the $x^+$ and $x^-$ directions, respectively.
} 
the equations of motion \eqref{eom:GWZW} become
\begin{subequations}
	\label{eom:GWZWgf}
	\begin{eqnarray}
	\label{eom:GWZWgfa}
	&\tg^{-1}\p_+ \tg=\mu\\
	\label{eom:GWZWgfb}
	&-\p_-\tg \tg^{-1}=\nu\\
	\label{eom:GWZWgfc}
	&-2\p_-\cbrac{\tg^{-1}\p_+ \tg}=0,
	\end{eqnarray}
\end{subequations}
respectively. In particular, \eqref{eom:GWZWgfa} and \eqref{eom:GWZWgfb} are the current constraints \eqref{eqn:current constraints(t)} and \eqref{eqn:current constraints(zbar)}. Furthermore, \eqref{eom:GWZWgfc} is the same equation of motion of the 
\textit{ungauged} 3d WZW model (see \eqref{eom:BC1t}). 
Following the steps seen in \S\ref{sec:3dToda}, we can then obtain the 3d Toda theory.

Therefore, the gauged 3d WZW model proposed here produces the same physics as the 3d Toda theory. We shall now attempt to obtain 3d analogues of W-algebras from this equivalent description of the 3d Toda theory, starting with a 3d analogue of the Virasoro algebra.

\bigskip\noindent\textit{Current Algebra of Gauged $SL(2,\complex)$ 3d WZW Model}
\vspace*{0.5em}

For clarity, we shall first focus on the simplest case of the gauged $SL(2,\complex)$ 3d WZW model that is equivalent to a 3d analogue of Liouville theory. In this case, $\sbrac{C_{ij}}=2$ and $\gamma_i=\half$. 
We shall compute Poisson brackets in the 3d gauged WZW model, where we shall take $x^-$ to be the temporal direction. We shall also write $x^+=\xi$ for presentation purposes. 

From \eqref{eqn:PB}, the symplectic form on the phase space is found to be
\begin{equation}
\omega=\half\int d\xi dz\,\tr\cbrac{\p_\xi\cbrac{\delta \tg(z,\xi)\tg^{-1}(z,\xi)}\wedge\delta \tg(z,\xi)\tg^{-1}(z,\xi)}.
\end{equation}
We moreover decompose the 3d WZW current in terms of basis generators of $SL(2,\complex)$, i.e.,
\begin{equation}
\label{eqn:PB current}
J_a=-{i\over 2} \tr\cbrac{\tau_a\tg^{-1}(z,\xi)\p_\xi \tg(z,\xi)},
\end{equation}
where $\tau_{a}$ denote $SL(2,\C)$ generators. 
Setting $\CO=J_a(z',\xi')$ and $\zeta=\tg(z,\xi)$ in \eqref{eqn:PB}, we obtain the Poisson bracket relations
\begin{equation}
\label{PB:Jg}
\PB{J_a(z',\xi'),\tg(z,\xi)}=-{i\over 2} \tg(z',\xi')\tau_{a}\delta(\xi-\xi')\delta(z-z').
\end{equation}
Likewise, setting $\CO=J_a(z',\xi')$ and $\zeta=\tg^{-1}(z,\xi)$ outputs
\begin{equation}
\label{PB:Jgi}
\PB{J_a(z',\xi'),\tg^{-1}(z,\xi)}={i\over 2} \tau_{a}\tg^{-1}(z',\xi')\delta(\xi-\xi')\delta(z-z').
\end{equation}
Next, using \eqref{PB:Jg} and \eqref{PB:Jgi}, the Poisson bracket relations for $J_a$ are obtained as
\begin{equation}
\label{PB:JJ}
\PB{J_a(z',\xi'),J_b(z,\xi)}=\cbrac{ i{\epsilon_{abc}}J^{c}(z,\xi)\delta(\xi-\xi')+\half\delta_{ab}\delta'(\xi-\xi')}\delta(z-z'),
\end{equation}
where $\delta'(\xi-\xi')=\p_\xi\delta(\xi-\xi')$. This takes the form of the ``analytically-continued" toroidal Lie algebra seen in \cite{ashwinkumar20203dwzw}.

\bigskip\noindent\textit{Virasoro Algebra of 3d Liouville Theory}
\vspace*{0.5em}

The next step is to derive the Poisson bracket relations for the 3d analogue of the chiral Sugawara energy-momentum tensor,
\begin{equation}
T(z,\xi)=J_a(z,\xi) J^a(z,\xi).
\end{equation}
Via \eqref{PB:JJ}, the Poisson bracket relations for $T(z,\xi)$ are obtained as
\begin{align}
\label{PB:TgJ}
&\PB{T(z',\xi'),J_a(z,\xi)}=\cbrac{\p_\xi J_a(z,\xi)\delta(\xi-\xi')+J_a(z,\xi)\delta'(\xi-\xi')}\delta(z-z')\\
\label{PB:TgTg}
&\PB{T(z',\xi'),T(z,\xi)}=\cbrac{\p_\xi T(z,\xi)\delta(\xi-\xi')+2T(z,\xi)\delta'(\xi-\xi')}\delta(z-z').
\end{align}

Next, we consider the Noether current given in \eqref{em:improved}. Via the Gauss decomposition (see \eqref{eqn:gauss decomposition}), and using the 3d Toda equations of motion \eqref{eom:3dtoda}, we obtain the relation
\begin{equation}
\label{eqn:Jphi}
J_3=-\half\p_\xi\phi.
\end{equation}
Thus, upon a rescaling, and using \eqref{eqn:Jphi}, we can also define \eqref{em:improved} as
\begin{equation}
\Theta(z,\xi)=T(z,\xi)-\p_\xi J_3(z,\xi).
\end{equation}
Finally, via \eqref{PB:TgJ} and \eqref{PB:TgTg}, we obtain
\begin{equation}
\label{PB:TheThe}
{\PB{\Theta(z',\xi'),\Theta(z,\xi)}=\left(\p_\xi\Theta(z,\xi)\delta(\xi-\xi')+2\Theta(z,\xi)\delta'(\xi-\xi')-\half\delta'''(\xi-\xi')\right)\delta(z-z')},
\end{equation}
which is a 3d analogue of the classical Virasoro algebra derived from the 3d Liouville theory.


In \cite{jorjadze2001poisson}, where the analogue of \eqref{PB:TheThe} is derived for the 2d Liouville theory,
Moyal quantization is shown to lead  
to its central extension, whereby one arrives at the well-known form of the standard Virasoro algebra, that can be equivalently obtained via canonical quantization.
We thus expect to obtain an analogous central extension to \eqref{PB:TheThe} via Moyal quantization, leading to an analogue of the standard, centrally extended Virasoro algebra. 
By analogy wth the 2d case, this algebra ought to take the form 
\begin{equation}
\label{com:TheThe}
    \boxed{\begin{aligned}
    \sbrac{\hTheta(z',\xi'),\hTheta(z,\xi)}=
    &\left(\vphantom{\half}\p_\xi\hTheta(z,\xi)\delta(\xi-\xi')+2\hTheta(z,\xi)\delta'(\xi-\xi')\right.\\
    &\quad\qquad\left.+\cbrac{k-\half\eta^2}\delta'''(\xi-\xi')
    +k\delta'(\xi-\xi')
    \right)\delta(z-z')
    \end{aligned}}
\end{equation}
where $k\in \real$ is a constant to be determined, and $\hat{\Theta}(z,\xi)=T(z,\xi)-\eta\p_\xi J_3(z,\xi)$,
where $\eta$ is a deformation parameter.

When we replace $SL(2,\complex)$ with $SL(N,\complex)$, we also expect to obtain a Virasoro algebra of the form \eqref{com:TheThe}. 
Moreover, we should be able to derive analogous results for higher spin currents in the 3d Toda theory via a generalized Sugawara construction, similar to how W-algebra currents are constructed in 2d Toda theory \cite{bouwknegt1993w}. In doing so, we ought to obtain analogues of W-algebras, similar in form to those that arise in 2d Toda theory, but with generators having holomorphic dependence on the Riemann surface, $\Sigma$, and a delta function $\delta(z-z')$ appearing as an overall factor on the RHS. 

Note that, so far, we have taken $x^-$ to be the temporal direction, and argued that this leads to a 3d analogue of a W-algebra via Moyal quantization. We could have equivalently taken $x^+$ to be the temporal direction, and doing so should lead to another copy of the W-algebra which is $x^-$-dependent instead.

\section{2d Twisted Sigma Model from Partially Twisted 5d MSYM\label{sec:4}}

\subsection{2d A-model on Bogomolny Moduli Space as an Effective Theory}\label{sec:gensigma}

Recall that we have started with the 5d ``GL''-twisted MSYM on $\CM=I\times\R_+\times S^1\times \Sigma$. We would now like to derive a 2d sigma model from our original setup, by taking the topological directions along $ I\times\R_+$ to be very large. 
To this end, we shall carry out the appropriate topological deformation of the metric, similar to the approach taken in \cite{bershadsky1995topological}. 

\bigskip\noindent\textit{Topological Deformation}
\vspace*{0.5em}



More explicitly, we shall consider a topological deformation of the form 
\begin{equation}
\label{eqn:topdef}
    \begin{aligned}
    ds^2&=\cbrac{dx^1}^2+\cbrac{dx^2}^2+\cbrac{dx^3}^2+\cbrac{dx^4}^2+\cbrac{dx^5}^2\\
    &\to\varepsilon^{-1}\cbrac{\cbrac{dx^1}^2+\cbrac{dx^2}^2}+\cbrac{dx^3}^2+\cbrac{dx^4}^2+\cbrac{dx^5}^2,
    \end{aligned}
\end{equation}
where $\varepsilon$ is taken to be very small. This deformation can be interpreted as making the volume of $I\times\R_+$ large, relative to the volume of $ S^1\times \Sigma$, which is equivalent to making the volume of $ S^1\times \Sigma$ infinitesimally small. This results in an effective 2d theory on $I\times\R_+$. 

Let us now examine the $A$- and $\varphi$- dependent part of the 5d action \eqref{action:5d MSYM}, which takes the form
\begin{equation}
    \label{action:5d A,phi}
    S_{\text{5d GL}}^{(A,\varphi)}=-{1\over {g_5}^2}\int_{\CM}d^5x\, \tr\cbrac{\half\CF_{\alpha\beta}\ov\CF^{\alpha\beta}+4\CF_{\alpha \zbar}{\ov\CF^{\alpha}}_{z}+\cbrac{D_{\alpha}\varphi^{\alpha}-2iF_{z\zbar}}^2},
\end{equation}
where $z,\zbar\in \Sigma$ and $\alpha,\beta=1,2,3$, and $\CF$ is the field strength corresponding to complex gauge fields $\CA_{\alpha}=A_{\alpha}+i\varphi_{\alpha}$, where we recall that $\CA_{z}=A_{z}$ and $\CA_{\zbar}=A_{\zbar}$. 

Following \cite{bershadsky1995topological}, we can decompose the fields as $A=A_{I\times\R_+}+A_{S^1}+A_{\Sigma}$ and $\varphi=\varphi_{I\times\R_+}+\varphi_{S^1}$, and rewrite \eqref{action:5d A,phi} as
\begin{equation}
    \label{action:5d A,phi(decomposed)}
    \begin{aligned}
    S_{\text{5d GL}}^{(A,\varphi)}=-{1\over {g_5}^2}\int_{\CM}d^5x\, \tr&
    \left(\half\CF_{\ta\tbe}\ov\CF^{\ta\tbe}
    +\CF_{\tau\ta}\ov\CF^{\tau\ta}
    +4\CF_{\tau \zbar}{\ov\CF^{\tau}}_{z}
    +4\CF_{\ta \zbar}{\ov\CF^{\ta}}_{z}\right.\\
    &\left.\quad+\cbrac{D_{\ta}\varphi^{\ta}}^2
    +2D_{\ta}\varphi^{\ta}\cbrac{D_{\tau}\varphi^{\tau}-2iF_{z\zbar}}
    +\cbrac{D_{\tau}\varphi^{\tau}-2iF_{z\zbar}}^2\right),
    \end{aligned}
\end{equation}
where $x^3=\tau$, and $\ta,\tbe=1,2$. 
Next, we carry out the deformation \eqref{eqn:topdef}, so that \eqref{action:5d A,phi(decomposed)} becomes
\begin{equation}
    \label{action:5d A,phi(deformed)}
    \begin{aligned}
    S_{\text{5d GL}}^{(A,\varphi)}=-{1\over {g_5}^2}\int_{\CM}d^5x\, \tr&
    \left(\half\varepsilon\CF_{\ta\tbe}\ov\CF^{\ta\tbe}
    +\CF_{\tau\ta}\ov\CF^{\tau\ta}
    +4\varepsilon^{-1}\CF_{\tau \zbar}{\ov\CF^{\tau}}_{z}
    +4\CF_{\ta \zbar}{\ov\CF^{\ta}}_{z}\right.\\
    &\left.\quad+\varepsilon\cbrac{D_{\ta}\varphi^{\ta}}^2
    +2D_{\ta}\varphi^{\ta}\cbrac{D_{\tau}\varphi^{\tau}-2iF_{z\zbar}}
    +\varepsilon^{-1}\cbrac{D_{\tau}\varphi^{\tau}-2iF_{z\zbar}}^2\right).
    \end{aligned}
\end{equation}
When we take the limit $\varepsilon\to 0$, terms proportional to $\varepsilon^{-1}$ will diverge. Hence, to ensure the finiteness of the action, it is necessary to impose the conditions
\begin{equation}
    \label{eom:hitchinlifted}
    \boxed{\begin{aligned}
    &F_{\tau\zbar}-iD_{\zbar}\varphi_\tau=0\\
    &D_{\tau}\varphi^{\tau}-2iF_{z\zbar}=0
    \end{aligned}}
\end{equation}
Interestingly, \eqref{eom:hitchinlifted} are the \textit{Bogomolny equations}, otherwise known as 3d monopole equations. Hence, solutions of \eqref{eom:hitchinlifted} span the moduli space of Bogomolny monopoles, which we shall denote as $\bogmoduli{ S^1\times \Sigma}$.

\bigskip\noindent\textit{2d Sigma Model Action from ``GL"-Twisted 5d MSYM}
\vspace*{0.5em}

Upon taking $\varepsilon\to 0$, and imposing the conditions \eqref{eom:hitchinlifted}, the remaining action becomes 
\begin{equation}
    \label{action:5d A,phi(remaining)}
    {S'}_{\text{5d GL}}^{(A,\varphi)}=-{1\over {g_5}^2}\int_{\CM}d^5x\, \tr
    \left(\CF_{ p \ta}\ov\CF^{ p \ta}\right),
\end{equation}
where $p=3,4,5$. 

In order to specify a solution of the Bogomolny equations, $A_{p}$ and $\varphi_{\tau}$, on $\CM$, we shall specify a map $X:I\times\R_+\to\bogmoduli{ S^1\times \Sigma}$. Then, we can write
\begin{subequations}
    \begin{eqnarray}
    &A_{ p }(x^1,x^2,x^3,x^4,x^5)=A_{ p }(x^3,x^4,x^5|X(x^1,x^2))\\
    &\varphi_{ \tau }(x^1,x^2,x^3,x^4,x^5)=\varphi_{ \tau }(x^3,x^4,x^5|X(x^1,x^2)).
    \end{eqnarray}
\end{subequations}
We can express the variations of the solutions of the Bogomolny equations, $\delta A_{ p }$ and $\delta\varphi_\tau$, in terms of cotangent vectors on $\bogmoduli{ S^1\times \Sigma}$, up to gauge transformations, i.e.,
\begin{subequations}
    \begin{eqnarray}
    &{\p A_{ p }\over\p X^I}=\delta_I A_{p }+D_{ p }E_I\\
    &{\p \varphi_{\tau}\over\p X^I}=\delta_I \varphi_{\tau}+[\varphi_\tau,E_I],
    \end{eqnarray}
\end{subequations}
where $I=1,\cdots,\text{dim}(\bogmoduli{ S^1\times \Sigma})$, and where $E$ is identified with a gauge connection on $\bogmoduli{ S^1\times \Sigma}$.

Next, we choose the gauge-fixing conditions $\ov{\mathcal{D}}^\tau(\delta_I A_{\tau}+i\delta_I \varphi_{\tau})=0$ and $D_z\delta_I A_{\zbar}=0$. Noting that $\CA_{\ta}$ becomes auxiliary in the $\varepsilon\to 0$ limit, we integrate it out by setting
\begin{equation}
    \CA_{\ta}=E_I\p_{\ta}X^{I},
\end{equation}
assuming that the operator $\ov{\mathcal{D}}_\tau{\mathcal{D}}^\tau+4D_zD_{\zbar}$ is invertible (this can be achieved via the inclusion of non-local operators on $I\times \R_+$) where $\mathcal{D}_\tau=D_\tau+i[\varphi_\tau,\cdot]$. 
Hence, the action \eqref{action:5d A,phi(remaining)} becomes
\begin{equation}
    \label{action:2d sigma (5d MSYM)}
    \boxed{\begin{aligned}
    {S'}^{(A,\varphi)}_{\text{5d GL}}&= -{1\over {g_5}^2}
    \int_{I\times\R_+}d^2x\,\cbrac{\int_{ S^1\times \Sigma}d^3x\,\tr
    \cbrac{{\delta_I A^p}\delta_J A_{p}+{\delta_I \varphi^{\tau}}\delta_J\varphi_{\tau}}}\p^{\ta}X^I\p_{\ta}X^J\\
    &=
    \int_{I\times\R_+}d^2x\,G^{\text{B}}_{IJ}\p^{\ta}X^I\p_{\ta}X^J
    \end{aligned}}
\end{equation}
where we have identified the metric on $\bogmoduli{ S^1\times \Sigma}$ as
\begin{equation}
{G}^{\text{B}}=-\frac{1}{g_5^2}\int_{ S^1\times \Sigma}d^3x\,\tr
    \cbrac{{\delta A}^{ p }\otimes\delta A_{ p }+\delta \vp^{\tau}\otimes\delta \vp_{\tau}}.
\end{equation}

The action \eqref{action:2d sigma (5d MSYM)} is indeed the action of a bosonic sigma model governing maps $X:I\times\R_+\to\bogmoduli{ S^1\times \Sigma}$. 
The full, partially twisted 5d gauge theory should likewise reduce in the same way to a topological sigma model on $I\times \R_+$. This sigma model is an A-model in a particular symplectic structure on $\bogmoduli{ S^1\times \Sigma}$. To see this, note that the non-$\CQ$-exact term
of the 5d gauge theory includes a term proportional to
	\begin{equation}
	\label{id:t-dependence}
	\begin{aligned}
 \int_{I\times\R_+\times S^1\times \Sigma } d^5x~\, \textrm{Tr}{\half\varepsilon^{\alpha\beta\gamma}}\cbrac{\half F_{\alpha 5}F_{\beta\gamma}-\partial_\alpha\cbrac{F_{\beta 4}\phi_\gamma}}.
\end{aligned}
	\end{equation}
In the reduction procedure we have outlined, this term reduces to an integral over a pullback to $I\times \R_+$ of a symplectic form on $\bogmoduli{ S^1\times \Sigma}$, that can be denoted as $\omega=\til{\Psi} \omega_K^B$, with 
\begin{equation}
    \boxed{\omega^{\text{B}}_{K} = \frac{1}{2\pi} \int_{ S^1\times \Sigma} d^3x~ \textrm{Tr} (\delta \vphi_{\tau}\wedge \delta A_4 - \delta A_5\wedge \delta A_{\tau})}
\end{equation}
where the subscript $K$ denotes that this is analogous to the symplectic structure $\omega_K$ on Hitchin's moduli space as defined in \cite{kapustin2006electric}, if we identify $A_5$ with $\vphi_4$ in the latter.\footnote{
In fact, upon dimensional reduction along $S^1\in\Sigma$ (for $\Sigma= \R\times S^1$ or $T^2$), we obtain the well-studied case where the target space is Hitchin's moduli space \cite{kapustin2006electric}.} This implies that the 2d sigma model is an A-model with symplectic structure $\til{\Psi}\omega^{\text{B}}_{K}$.
In fact, the  entire non-$\CQ$-exact sector in \eqref{action00} reduces as
\begin{equation}
    \frac{\til{\Psi}}{4\pi}\int_{I\times\R_+\times S^1\times \Sigma}\ dz\wedge\tr\cbrac{\cF \wedge \cF}\rightarrow \int_{I\times \R_+} X^*(\omega-iB),
\end{equation}
where $B=-\til{\Psi}\omega_I^{\text{B}}$ is a $B$-field, with 
\begin{equation}
    \omega^{\text{B}}_{I} = -\frac{1}{2\pi} \int_{ S^1\times \Sigma} d^3x~ \textrm{Tr} (\delta A_{\tau}\wedge \delta A_4 - \delta \vp_{\tau} \wedge \delta A_{5}),
\end{equation}
which is the analogue of the symplectic structure $\omega_I$ on Hitchin's moduli space.

Note that, upon including in the 5d ``GL"-twisted theory non-local operators along $I\times \R_+$ that we expect to correspond 
to local operators in 3d Toda theory, the sigma model we obtain will include non-local operators on the worldsheet, while the target space will have dependence on the points on $S^1\times \Sigma$ where the operators are located.

\subsection{Physical States of 2d Sigma Model}

\begin{figure}[t!]
	\centering
\tikzset{every picture/.style={line width=0.75pt}} 
\begin{tikzpicture}[x=0.75pt,y=0.75pt,yscale=-1,xscale=1]
\draw    (60,83) -- (60.16,210.2) ;
\draw    (156.94,83) -- (157.1,210.2) ;
\draw    (60.16,214.2) -- (60.16,221.2) ;
\draw    (60.16,225.2) -- (60.16,232.2) ;
\draw    (60.16,236.2) -- (60.16,243.2) ;
\draw    (157.1,215.2) -- (157.1,222.2) ;
\draw    (157.1,226.2) -- (157.1,233.2) ;
\draw    (157.1,237.2) -- (157.1,244.2) ;
\draw    (156.94,83) -- (60,83) ;
\draw    (212.01,179.45) -- (260.01,179.45) ;
\draw [shift={(262.01,179.45)}, rotate = 180] [color={rgb, 255:red, 0; green, 0; blue, 0 }  ][line width=0.75]    (10.93,-3.29) .. controls (6.95,-1.4) and (3.31,-0.3) .. (0,0) .. controls (3.31,0.3) and (6.95,1.4) .. (10.93,3.29)   ;
\draw    (560.1,215.2) -- (560.1,222.2) ;
\draw    (560.1,226.2) -- (560.1,233.2) ;
\draw    (560.1,237.2) -- (560.1,244.2) ;
\draw    (560.01,112.45) .. controls (566.01,135.45) and (562.01,160.45) .. (580.01,159.45) ;
\draw    (530.1,80.51) -- (531.55,87.36) ;
\draw    (532.38,91.28) -- (533.82,98.12) ;
\draw    (534.65,102.04) -- (536.1,108.89) ;
\draw    (552.99,80.54) -- (554.49,87.37) ;
\draw    (555.35,91.28) -- (556.85,98.12) ;
\draw    (557.71,102.03) -- (559.21,108.86) ;
\draw    (302.16,213.2) -- (302.16,220.2) ;
\draw    (302.16,224.2) -- (302.16,231.2) ;
\draw    (302.16,235.2) -- (302.16,242.2) ;
\draw    (399.1,214.2) -- (399.1,221.2) ;
\draw    (399.1,225.2) -- (399.1,232.2) ;
\draw    (399.1,236.2) -- (399.1,243.2) ;
\draw    (302.16,209.2) .. controls (301.01,141.45) and (326.01,102.45) .. (286.01,83.45) ;
\draw    (399.1,209.2) .. controls (393.01,117.45) and (379.01,105.45) .. (412.01,83.45) ;
\draw    (286.01,83.45) .. controls (323.01,101.45) and (372.01,113.45) .. (412.01,83.45) ;
\draw    (560.35,211.09) .. controls (565.35,185.09) and (546.01,165.45) .. (537.1,111.2) ;
\draw    (600.38,215.2) -- (600.38,222.2) ;
\draw    (600.38,226.2) -- (600.38,233.2) ;
\draw    (600.38,237.2) -- (600.38,244.2) ;
\draw    (600.45,112.45) .. controls (594.32,135.45) and (598.4,160.45) .. (580.01,159.45) ;
\draw    (631.01,80.51) -- (629.53,87.36) ;
\draw    (628.69,91.28) -- (627.21,98.12) ;
\draw    (626.36,102.04) -- (624.88,108.89) ;
\draw    (607.63,80.54) -- (606.09,87.37) ;
\draw    (605.21,91.28) -- (603.68,98.12) ;
\draw    (602.8,102.03) -- (601.26,108.86) ;
\draw    (600.35,212.09) .. controls (595.24,186.09) and (614.75,165.45) .. (623.86,111.2) ;
\draw    (435.01,179.45) -- (483.01,179.45) ;
\draw [shift={(485.01,179.45)}, rotate = 180] [color={rgb, 255:red, 0; green, 0; blue, 0 }  ][line width=0.75]    (10.93,-3.29) .. controls (6.95,-1.4) and (3.31,-0.3) .. (0,0) .. controls (3.31,0.3) and (6.95,1.4) .. (10.93,3.29)   ;
\draw (40,156) node  [align=left] {$\displaystyle \mathcal{B}_{\text{Op}}$};
\draw (180,156) node  [align=left] {$\displaystyle \mathcal{B} '_{\text{Op}}$};
\draw (107,68) node  [align=left] {$\displaystyle \mathcal{B}_{\text{c}}$};
\draw (582,137) node  [align=left] {$\displaystyle \mathcal{B}_{\text{c}}$};
\draw (412,137) node  [align=left] {$\displaystyle \mathcal{B} '_{\text{Op}}$};
\draw (624,190) node  [align=left] {$\displaystyle \mathcal{B} '_{\text{Op}}$};
\draw (289,127) node  [align=left] {$\displaystyle \mathcal{B}_{\text{Op}}$};
\draw (538,186) node  [align=left] {$\displaystyle \mathcal{B}_{\text{Op}}$};
\draw (354,86) node  [align=left] {$\displaystyle \mathcal{B}_{\text{c}}$};
\end{tikzpicture}
\caption{Topological deformation of sigma model worldsheet $I\times\R_+$.}
	\label{fig:deformed worldsheet}
\end{figure}
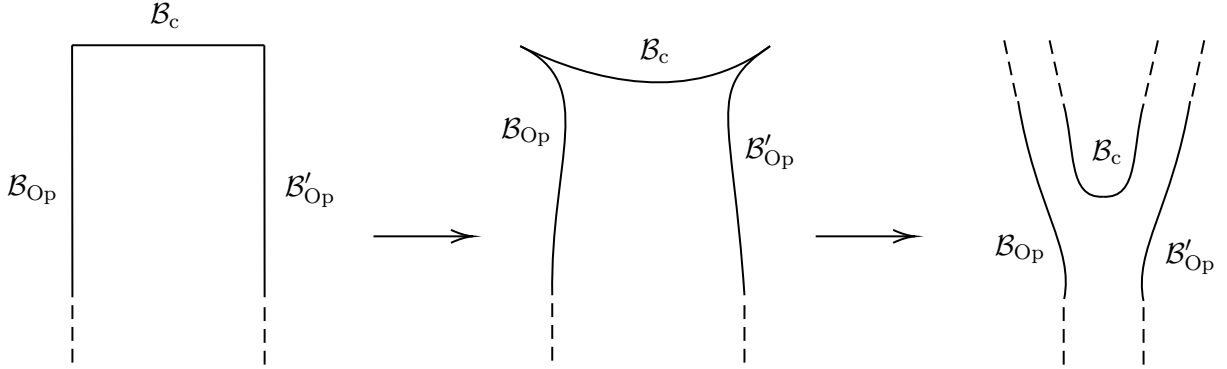

The NS5-type boundary condition of the 5d gauge theory gives rise to a space-filling coisotropic A-brane, $\Bc$, of the sigma model on $\bogmoduli{ S^1\times \Sigma}$. Defining complex and symplectic structures analogous to those on Hitchin's moduli space (identifying $A_5$ with $\vphi_4$ as before), this is a $(B,A,A)$ brane. To see this, we first note that the NS5-type boundary condition implies Neumann boundary conditions on $\CA_{\zbar}$ and $\CA_{\tau}$, so the corresponding brane must be space-filling. 
Secondly, 
 $\omega$ and $B$ satisfy $(\omega^{-1} B)^2=-1$, which implies that the brane is an A-brane in symplectic structure $\omega=\til{\Psi} \omega_K^{\text{B}}$. Finally, the $B$-field is of type $(1,1)$ in complex structure $I^{\text{B}}$, implying that the brane is a B-brane in this complex structure.

On the other hand, the Nahm pole-type boundary conditions of the 5d gauge theory give rise to
Lagrangian branes $\Bop$ and $\Bop'$ in $\bogmoduli{ S^1\times \Sigma}$. These are
analogues of branes of opers in Hitchin's moduli space \cite{nekrasov2010omega}, and are both $(A,B, A)$ branes, since $\omega_I^{\text{B}}$ and $\omega_K^{\text{B}}$ both vanish on the support of these branes, and are maximal with this property.

It can be shown that the Lagrangian branes $\Bop$ and $\Bop'$ only intersect at one point, i.e., the origin. This implies that the space of states of the $(\Bop,\Bop')$ string contains only a single state.
To find more states, we can topologically deform the worldsheet $I\times\real_+$ by pinching off the corners to infinity to form a Y-shaped worldsheet (see Figure \ref{fig:deformed worldsheet}). This gives rise to $(\Bc,\Bop)$ and $(\Bc,\Bop')$ strings that originate from the corners of $I\times\real_+$. The physical space of states of each type of string corresponds to a space of $J^{\text{B}}$-holomorphic sections of a bundle on the corresponding Lagrangian brane. Specifically, the two spaces of states can be computed (following \cite{kapustin2006electric}) to be
\begin{subequations}
\begin{eqnarray}
&\mathcal{H}_{(\Bc,\Bop)}&=H^0(\Bop,K^{1/2}_{\Bop})\\
&\mathcal{H}_{(\Bc,\Bopp)}&=H^0(\Bopp,K^{1/2}_{\Bopp}),
\end{eqnarray}
\end{subequations}
where $K_{\Bop}$ is the canonical line bundle over the brane $\Bop$.

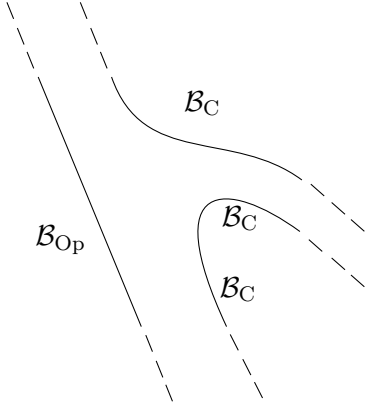
\begin{figure}[t!]
    \centering
\scalebox{1}{\begin{tikzpicture}[x=0.75pt,y=0.75pt,yscale=-1,xscale=1]
\draw    (56.68,62.21) -- (108.68,188.21) ;
\draw    (93.68,62.21) .. controls (109.68,108.21) and (154.68,88.21) .. (189.68,114.21) ;
\draw    (151.68,188.21) .. controls (116.68,113.21) and (151.68,113.21) .. (188.68,140.21) ;
\draw    (110.95,192.51) -- (114.31,200.86) ;
\draw    (116.18,205.5) -- (119.55,213.85) ;
\draw    (121.04,217.56) -- (124.4,225.9) ;
\draw    (153.52,192.16) -- (157.6,200.19) ;
\draw    (159.86,204.64) -- (163.94,212.67) ;
\draw    (165.76,216.23) -- (169.83,224.25) ;
\draw    (40.91,24.53) -- (44.29,32.87) ;
\draw    (46.17,37.5) -- (49.56,45.84) ;
\draw    (51.06,49.55) -- (54.44,57.89) ;
\draw    (77.91,24.53) -- (81.29,32.87) ;
\draw    (83.17,37.5) -- (86.56,45.84) ;
\draw    (88.06,49.55) -- (91.44,57.89) ;
\draw    (194.18,116.3) -- (200.93,122.26) ;
\draw    (204.68,125.56) -- (211.43,131.51) ;
\draw    (214.43,134.16) -- (221.18,140.11) ;
\draw    (193.18,144.3) -- (199.93,150.26) ;
\draw    (204.68,153.56) -- (211.43,159.51) ;
\draw    (214.43,162.16) -- (221.18,168.11) ;
\draw (147,162) node [anchor=north west][inner sep=0.75pt]   [align=left] {$\displaystyle \mathcal{B}_{\text{C}}$};
\draw (148,126) node [anchor=north west][inner sep=0.75pt]   [align=left] {$\displaystyle \mathcal{B}_{\text{C}}$};
\draw (129,68) node [anchor=north west][inner sep=0.75pt]   [align=left] {$\displaystyle \mathcal{B}_{\text{C}}$};
\draw (54,136) node [anchor=north west][inner sep=0.75pt]   [align=left] {$\displaystyle \mathcal{B}_{\text{Op}}$};
\end{tikzpicture}}
    \caption{The action of the algebra of $(\Bc,\Bc)$ strings on the space of states of $(\Bc,\Boper)$ can be realized by attaching the $(\Bc,\Bc)$ string to the end of the $(\Bc,\Boper)$ string.}
    \label{fig:actionstring}
\end{figure}

Moreover, the algebra of ($\Bc,\Bc$) strings acts on the space of states of ($\Bc,\Bop$) and ($\Bc,\Bopp$) strings, by attaching to the appropriate ends of the latter (see Figure \ref{fig:actionstring}). 
This is just the quantized algebra of $J^{\text{B}}$-holomorphic functions on $\bogmoduli{ S^1\times \Sigma}$ \cite{kapustin2006electric}, which are generated by $\cA_{+}$ and $\cA_{-}$. The algebra of ($\Bop,\Bop$) and ($\Bop',\Bop'$) strings likewise act on $\mathcal{H}_{(\Bc,\Bop)}$ and $\mathcal{H}_{(\Bc,\Bop')}$ respectively. These are expected to be the same quantized algebra, albeit with different deformation parameters.

\section{The 3d-2d Correspondence}\label{sec:5}

\begin{figure}[h!]
	\centering
	\tikzset{every picture/.style={line width=0.75pt}} 
	\begin{tikzpicture}[x=0.75pt,y=0.75pt,yscale=-1,xscale=1]

\draw   (206.32,258.11) .. controls (198.08,248.14) and (189.8,242.06) .. (181.51,239.87) .. controls (189.82,238.17) and (198.18,232.59) .. (206.55,223.12) ;
\draw    (205,235) -- (289,235) ;
\draw    (205,246) -- (289,246) ;
\draw    (229.51,72.76) -- (152.39,100.09) ;
\draw [shift={(150.51,100.76)}, rotate = 340.48] [color={rgb, 255:red, 0; green, 0; blue, 0 }  ][line width=0.75]    (10.93,-3.29) .. controls (6.95,-1.4) and (3.31,-0.3) .. (0,0) .. controls (3.31,0.3) and (6.95,1.4) .. (10.93,3.29)   ;
\draw    (144,171) -- (144,203.2) ;
\draw [shift={(144,205.2)}, rotate = 269.99] [color={rgb, 255:red, 0; green, 0; blue, 0 }  ][line width=0.75]    (10.93,-3.29) .. controls (6.95,-1.4) and (3.31,-0.3) .. (0,0) .. controls (3.31,0.3) and (6.95,1.4) .. (10.93,3.29)   ;
\draw    (318.68,73.21) -- (411.37,180.69) ;
\draw [shift={(412.68,182.21)}, rotate = 229.23] [color={rgb, 255:red, 0; green, 0; blue, 0 }  ][line width=0.75]    (10.93,-3.29) .. controls (6.95,-1.4) and (3.31,-0.3) .. (0,0) .. controls (3.31,0.3) and (6.95,1.4) .. (10.93,3.29)   ;
\draw   (288.51,257.09) .. controls (296.69,247.61) and (304.68,242.01) .. (312.45,240.3) .. controls (304.88,238.13) and (297.51,232.07) .. (290.35,222.14) ;

\draw    (214.95,3.66) -- (338.95,3.66) -- (338.95,72.66) -- (214.95,72.66) -- cycle  ;
\draw (276.95,38.16) node   [align=center] {
5d MSYM\\on\\$\displaystyle I  \times \mathbb{R}_+ \times S^1  \times \Sigma$
};
\draw    (102.95,102.66) -- (184.95,102.66) -- (184.95,171.66) -- (102.95,171.66) -- cycle  ;
\draw (143.95,137.16) node   [align=center] {
4d CS\\on\\$\displaystyle I\times  S^1 \times \Sigma$
};
\draw    (113.45,206.66) -- (174.45,206.66) -- (174.45,275.66) -- (113.45,275.66) -- cycle  ;
\draw (143.95,241.16) node   [align=center] {
3d Toda\\on\\$\displaystyle  S^1 \times \Sigma$
};
\draw    (320.45,185.66) -- (513.45,185.66) -- (513.45,282.66) -- (320.45,282.66) -- cycle  ;
\draw (416.95,234.16) node   [align=center] {
2d A-model\\on\\$\displaystyle I\times \mathbb{R}_+$\\with target $\bogmoduli{ S^1\times \Sigma}$
};
\draw (195,85.2) node [anchor=north west][inner sep=0.75pt]    {$\partial _{\mathbb{R}_+}$};
\draw (356,96) node [anchor=north west][inner sep=0.75pt]    {Scaling up $ I\times\R_+$};
\draw (146,176) node [anchor=north west][inner sep=0.75pt]    {${I}$};
\end{tikzpicture}
	\caption{Outline of the steps taken in this paper, repeated here for brevity.}
	\label{fig:5d}
\end{figure}
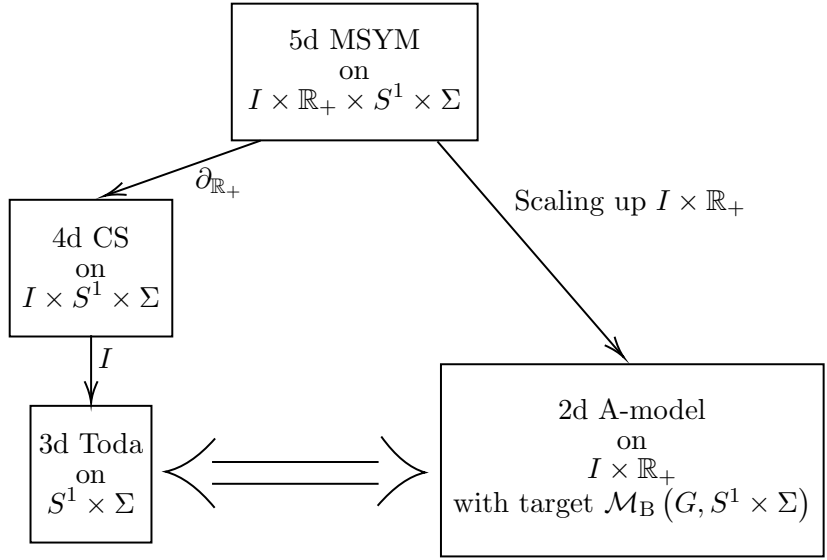

The 5d ``GL''-twisted MSYM on $I\times\R_+\times S^1\times \Sigma$, with NS5-type boundary conditions on $\real_+$ and Nahm pole-type boundary conditions on $I$, has given rise to two different effective descriptions.

In \S\ref{sec:3}, we localized ``GL-twisted" 5d MSYM to the 4d CS theory at the origin of $\real_+$, which in turn, is equivalent to 3d WZW models at the boundaries of $I$. Further implementing relevant constraints that originate from the Nahm pole-type boundary conditions (namely \eqref{con:nahm0} and \eqref{con:nahmpi}), the 3d WZW models are then identified with 3d Toda theory on $ S^1\times \Sigma$.

Via a gauged version of the 3d WZW model, we also argued for the existence of 3d analogues of W-algebras, for which the corresponding modules must be identified with 3d Toda states. There are two copies of such a W-algebra, corresponding to a choice of either $x^+$ or $x^-$ as the temporal direction.

In \S\ref{sec:4}, we argued that 
the 5d ``GL''-twisted theory gives rise, upon scaling up $ I \times \R_+$, to a 2d A-model of maps $X:I\times\R_+\to\bogmoduli{ S^1\times \Sigma}$.  Furthermore, via a topological deformation of the worldsheet $I\times\real_+$ to a Y-shaped configuration, we also computed the physical states of the A-model.


Since the $\CQ$-cohomology of states of the 5d theory ought to remain invariant in reducing to these two effective descriptions, we may identify 3d Toda states with physical states of the 2d A-model. 
Hence, we establish a \textit{novel} correspondence between the 3d Toda theory and a topological sigma model with A-branes, i.e., we have
\begin{equation}
\boxed{\begin{aligned}
	&\text{3d Toda theory}\\ 
	&\quad\qquad\text{on}\\
	&\qquad S^1\times \Sigma
	\end{aligned}\qquad\Huge{\text{$\Leftrightarrow$}}\normalsize\text{}\qquad
	\begin{aligned}
	&\text{Topological sigma model with A-branes}\\ 
	&\quad\qquad\qquad\qquad\text{on}\\
	&\qquad\qquad\qquad I\times \R_+\\
	&\qquad\text{with target }\bogmoduli{ S^1\times \Sigma }
	\end{aligned}}
\end{equation}
Mathematically, this implies that 
\begin{equation}
\boxed{\begin{aligned}
	&\text{Modules of 3d W-algebras}\\ 
	&\qquad\quad\text{defined on}\\
	&\qquad\qquad S^1\times \Sigma
	\end{aligned}\quad\Huge{\text{$\Leftrightarrow$}}\normalsize\text{}\quad
	\begin{aligned}
	&\text{$H^0(\Bop,K^{1/2}_{\Bop})\otimes H^0(\Bopp,K^{1/2}_{\Bopp})$}\\ 
	\end{aligned}}
\end{equation}
In other words,
\begin{equation*}
    \boxed{\begin{aligned}
    &\quad\text{Modules of 3d W-algebras defined on $ S^1\times \Sigma$ are modules for the quantized algebra of}\quad\\
    &\quad\text{$J^{\text{B}}$-holomorphic functions on $\bogmoduli{ S^1\times \Sigma }$.}
    \end{aligned}}
\end{equation*}
\nn
The steps taken in arriving at these results are summarized in Figure \ref{fig:5d}.

A direct approach to understanding the 3d-2d correspondence described above is as follows.  The 2d A-model action is a sum of a $\CQ$-exact term and a non-$\CQ$-exact term that can be written as 
\begin{equation}
 -\frac{\tilde{\Psi}}{4\pi}X^*\bigg(\int_{S^1\times \Sigma} dz \wedge \delta \mathcal{A}\wedge \delta \mathcal{A}\bigg)=\tilde{\Psi}X^*\Omega_J^B,   
\end{equation}
 where $\Omega_J^B$ is the the analogue of the complex symplectic form  $\Omega_J=-\frac{i}{4\pi} \int_C \textrm{Tr} \delta \mathcal{A}\wedge \delta \mathcal{A}$ on Hitchin's moduli space $\mathcal{M}_H(G,C)$, understood as the moduli space of flat $G_{\mathbb{C}}$-connections on $C$. Since $\Omega_J$ is the natural symplectic form with respect to which 3d analytically-continued Chern-Simons theory can be quantized via geometric quantization, one can similarly identify  $\Omega^B_J$ as the relevant symplectic form for the quantization of 4d Chern-Simons theory. Now, via the topological symmetry of the 2d A-model, one can shrink the $I\times \mathbb{R}_+$ worldsheet to the half-line $\mathbb{R}_+$, thereby implementing the constraints coming from the Nahm pole-type boundary conditions into $\Omega^B_J$. The resulting symplectic form can then be identified with a symplectic form on the phase space of 3d Toda theory, and in the 1d description on $\mathbb{R}_+$ reduces to a symplectic potential via Stokes' theorem. Localization of the quantum mechanical model means that the $\CQ$-exact term drops out, and one is left with a path integral describing quantum mechanics on the phase space of 3d Toda theory. Thus, the 3d-2d correspondence is essentially the identification of quantum 3d Toda theory with the quantization of the phase space of classical 3d Toda theory.
\section{Conclusion and Future Work}

In this work, we have derived a 3d analogue of Toda theory, and shown that it is dual to a 2d A-model on Bogomolny moduli space. Moreover, we have found that modules of 3d W-algebras are modules for a quantized algebra of holomorphic functions on the Bogomolny moduli space. The crucial ingredient is the fact that the 5d $\mathcal{N}=2$ SYM theory  admits a partial twist analogous to the 4d GL-twist, that is topological-holomorphic.

This suggests a generalization. Namely, it will be interesting to define a GL-type partial twist (in two directions) for maximally supersymmetric Yang-Mills theory in 6d, whereby one expects to relate 4d analogues of W-algebras to a quantized algebra of certain holomorphic functions on the moduli space of instantons. Such 4d W-algebras should arise from a 4d analogue of Toda theory, that should be the boundary dual of 5d Chern-Simons theory with Nahm pole-type boundary conditions.

In addition, it would be interesting to embed the 5d ``GL''-twisted theory in a partial twist of the 6d $\mathcal{N}=(2,0)$ SCFT. This is likely to lead to a 3d-3d duality involving the 3d Toda theory we have derived in this work.

We hope to explore these ideas in future work.

\appendix

\section{Poisson Brackets}
\label{append:symplectic}
\bigskip\noindent\textit{Derivation of Poisson Brackets from Symplectic Form}
\vspace*{0.5em}

Generically, the Poisson brackets of two operators can be deduced from a non-degenerate symplectic form \cite{jorjadze2001poisson,witten1984non} $\omega=\omega_{ab}(\zeta)\,\delta\zeta^a\wedge\delta\zeta^b$ via
\begin{equation}
-\delta\CO=\omega_{ab}(\zeta)\delta\zeta^a\PB{\CO,\zeta^b},
\end{equation}
where $\delta$ denotes a variation. Then, the Poisson bracket can be deduced by comparing the symplectic form with the variation of some operator $\CO$.

In the continuous case, where $\omega=\int dx \omega(x)\delta\zeta(x)\wedge\delta\zeta(x)$, this becomes
\begin{equation}
\label{eqn:PB}
-\delta\CO(x)=\int dy\, \omega(y)\delta\zeta(y)\PB{\CO(x),\zeta(y)}.
\end{equation}
In the same manner, the Poisson bracket can be deduced by comparing the symplectic form with the variation of some operator $\CO(x)$.

\bibliographystyle{ieeetr}
\bibliography{bib}

\begin{thebibliography}{10}

\bibitem{alday2010liouville}
L.~Alday, D.~Gaiotto, and Y.~Tachikawa, ``{Liouville Correlation Functions from
  Four-dimensional Gauge Theories},'' {\em Letters in Mathematical Physics},
  vol.~91, no.~2, pp.~167--197, 2010.
\newblock
  [\href[pdfnewwindow=true]{https://arxiv.org/abs/0906.3219}{arXiv:0906.3219}].

\bibitem{wyllard20091}
N.~Wyllard, ``{$A_{N- 1}$ conformal Toda field theory correlation functions
  from conformal \susy{2} $SU(N)$ quiver gauge theories},'' {\em Journal of
  High Energy Physics}, vol.~2009, no.~11, p.~002, 2009.
\newblock
  [\href[pdfnewwindow=true]{https://arxiv.org/abs/0907.2189}{arXiv:0907.2189}].

\bibitem{nekrasov2010omega}
N.~Nekrasov and E.~Witten, ``{The Omega Deformation, Branes, Integrability and
  Liouville Theory},'' {\em Journal of High Energy Physics}, vol.~2010, no.~9,
  p.~92, 2010.
\newblock
  [\href[pdfnewwindow=true]{https://arxiv.org/abs/1002.0888}{arXiv:1002.0888}].

\bibitem{costello2017gaugeI}
K.~Costello, E.~Witten, and M.~Yamazaki, ``{Gauge Theory And Integrability,
  I},'' in {\em Notices of the International Congress of Chinese
  Mathematicians}, vol.~6, pp.~46--119, International Press of Boston, 2018.
\newblock
  [\href[pdfnewwindow=true]{https://arxiv.org/abs/1709.09993}{arXiv:1709.09993}].

\bibitem{costello2018gaugeII}
K.~Costello, E.~Witten, and M.~Yamazaki, ``{Gauge Theory And Integrability,
  II},'' in {\em Notices of the International Congress of Chinese
  Mathematicians}, vol.~6, pp.~120--146, International Press of Boston, 2018.
\newblock
  [\href[pdfnewwindow=true]{https://arxiv.org/abs/1802.01579}{arXiv:1802.01579}].

\bibitem{ashwinkumar2019unifying}
M.~Ashwinkumar and M.-C. Tan, ``{Unifying Lattice Models, Links and Quantum
  Geometric Langlands via Branes in String Theory},'' {\em {\normalfont to
  appear in} Advances in Theoretical and Mathematical Physics}.
\newblock
  [\href[pdfnewwindow=true]{https://arxiv.org/abs/1910.01134}{arXiv:1910.01134}].

\bibitem{geyer2003higher}
B.~Geyer and D.~M{\"u}lsch, ``{Higher-dimensional analogue of the
  {Blau}--{Thompson} model and $N_T= 8$, $D= 2$ Hodge-type cohomological gauge
  theories},'' {\em Nuclear Physics B}, vol.~662, no.~3, pp.~531--553, 2003.
\newblock
  [\href[pdfnewwindow=true]{https://arxiv.org/abs/hep-th/0211061}{hep-th/0211061}].

\bibitem{cordova2017five}
C.~C{\'o}rdova and D.~L. Jafferis, ``Five-dimensional maximally supersymmetric
  {Yang}-{Mills} in supergravity backgrounds,'' {\em Journal of High Energy
  Physics}, vol.~2017, no.~10, p.~3, 2017.
\newblock
  [\href[pdfnewwindow=true]{https://arxiv.org/abs/1305.2886}{arXiv:1305.2886}].

\bibitem{kapustin2006electric}
A.~Kapustin and E.~Witten, ``{Electric-Magnetic Duality And The Geometric
  Langlands Program},'' {\em Communications in Number Theory and Physics},
  vol.~1, no.~1, pp.~1--236, 2007.
\newblock
  [\href[pdfnewwindow=true]{https://arxiv.org/abs/hep-th/0604151}{hep-th/0604151}].

\bibitem{elliott2019multiplicative}
C.~Elliott and V.~Pestun, ``{Multiplicative Hitchin Systems and Supersymmetric
  Gauge Theory},'' {\em Selecta Mathematica}, vol.~25, no.~4, p.~64, 2019.
\newblock
  [\href[pdfnewwindow=true]{https://arxiv.org/abs/1812.05516}{arXiv:1812.05516}].

\bibitem{mikhaylov2018teichmuller}
V.~Mikhaylov, ``{Teichm{\"u}ller TQFT vs. Chern-Simons Theory},'' {\em Journal
  of High Energy Physics}, vol.~2018, no.~4, p.~85, 2018.
\newblock
  [\href[pdfnewwindow=true]{https://arxiv.org/abs/1710.04354}{arXiv:1710.04354}].

\bibitem{gaiotto2012knot}
D.~Gaiotto and E.~Witten, ``{Knot Invariants from Four-Dimensional Gauge
  Theory},'' {\em Advances in Theoretical and Mathematical Physics}, vol.~16,
  no.~3, pp.~935--1086, 2012.
\newblock
  [\href[pdfnewwindow=true]{https://arxiv.org/abs/1106.4789}{arXiv:1106.4789}].

\bibitem{ashwinkumar20203dwzw}
M.~Ashwinkumar, ``{Integrable Lattice Models and Holography},'' {\em Journal of
  High Energy Physics}, no.~2, p.~227, 2021.
\newblock
  [\href[pdfnewwindow=true]{https://arxiv.org/abs/2003.08931}{arXiv:2003.08931}].

\bibitem{jorjadze2001poisson}
G.~Jorjadze and G.~Weigt, ``{Poisson Structure and Moyal Quantisation of the
  Liouville Theory},'' {\em Nuclear Physics B}, vol.~619, no.~1-3,
  pp.~232--256, 2001.
\newblock
  [\href[pdfnewwindow=true]{https://arxiv.org/abs/hep-th/0105306}{hep-th/0105306}].

\bibitem{bouwknegt1993w}
P.~Bouwknegt and K.~Schoutens, ``{W-symmetry in Conformal Field Theory},'' {\em
  Physics Reports}, vol.~223, no.~4, pp.~183--276, 1993.
\newblock
  [\href[pdfnewwindow=true]{https://arxiv.org/abs/hep-th/9210010}{hep-th/9210010}].

\bibitem{bershadsky1995topological}
M.~Bershadsky, A.~Johansen, V.~Sadov, and C.~Vafa, ``{Topological Reduction of
  4D SYM to 2D $\sigma$-Models},'' {\em Nuclear Physics B}, vol.~448, no.~1-2,
  pp.~166--186, 1995.
\newblock
  [\href[pdfnewwindow=true]{https://arxiv.org/abs/hep-th/9501096}{hep-th/9501096}].

\bibitem{witten1984non}
E.~Witten, ``{Non-abelian bosonization in two dimensions},'' {\em
  Communications in mathematical physics}, vol.~92, no.~4, pp.~455--472, 1984.

\end{thebibliography}
\end{document}